\documentclass{aa}  
\usepackage{graphicx}
\usepackage{txfonts}
\begin{document}
   \title{Double Relics in Abell 2345 and Abell 1240}
   \subtitle{spectral index and polarization analysis}
   \author{A. Bonafede
          \inst{1,2}
	  \and G. Giovannini\inst{1,2}
	  \and L. Feretti\inst{2}
	  \and F. Govoni\inst{3}
	   \and M. Murgia\inst{2,3}
          }
         
   \offprints{bonafede@ira.inaf.it}
   \institute{Universit\'a di Bologna, Dip. di Astronomia, via Ranzani 1, I-40126 Bologna, Italy\\
         \and
             INAF, Istituto di Radioastronomia, via Gobetti 101, I-40129 Bologna, Italy\\
	 \and
	     INAF, Osservatorio Astronomico di Cagliari, Strada 54, Loc. Poggio dei Pini, I-09012 Capoterra (Ca), Italy\\
             }
   \date{Received July 14, 2008; accepted September 19, 2008}
% \abstract{}{}{}{}{} 
% 5 {} token are mandatory
  \abstract 
% context heading (optional) % {} leave it empty inecessary 
{} 
% aims heading (mandatory) 
  {The aim of the present work is to study the radio
   properties of double relics in Abell 1240 and Abell 2345 in the 
   framework of double relic formation models.}
%   methods heading (mandatory) 
  {We present new Very Large Array observations at 20 and
   90 cm for these two clusters.  We performed spectral index and
   polarization analysis and compared our results with expectations from
   theoretical models.}
%   results heading (mandatory) 
{The presence of double relics in these two cluster is confirmed by
these new observations. Double relics in Abell 1240 show radio
morphology, spectral index and polarization values in agreement with
``outgoing merger shocks'' models. One of the relics of Abell 2345,
shows a peculiar morphology and spectral index profile, that are
difficult to reconcile with present scenarios. We suggest a possible
origin for this peculiar relic.}
% conclusions heading   (optional), leave it empty if necessary 
{} \keywords{Galaxies:clusters:general -
   Galaxies:clusters:individual:A1240, A2345 - Radiation mechanism:non-thermal - 
  Polarization - Magnetic field} 
\maketitle
%
%________________________________________________________________

\section{Introduction}
The hierarchical model of structure formation predicts that clusters
of galaxies form by subsequent merging of smaller structures. A
consistent amount of energy ($\sim 10^{63} - 10^{64}$ ergs) is
released in the Intra Cluster Medium (ICM) as result of such merger
events. Cosmological numerical simulations have shown that shocks and
turbulence associated with these processes do not only heat the ICM
but also play an important role in non-thermal phenomena occurring in
the ICM (see e. g. the review by Dolag et al. 2008). Although these
processes are not completely understood yet, this energy input is
supposed to accelerate and inject relativistic particles on cluster
scale and amplify the magnetic field in the ICM.\\ Extended radio
sources on cluster scale not associated with any optical counterpart
but arising from the ICM have been detected in an increasing number of
galaxy clusters. They are called radio halos and radio relics
depending on their morphology, location and radio properties. \\ Radio
relics are extended radio sources located at the outskirts of galaxy
clusters, and are strongly polarized, with linear fractional
polarization at 20 cm above 10 $\%$, reaching values up to 50 $\%$ in
some regions (see e.g. Govoni \& Feretti 2004; Ferrari et
al. 2008). Their origin is debated and not well known.  There is a
general consensus that it is related to phenomena occurring in the ICM
during merging events. So far, there are $\sim$ 20 clusters of
galaxies where at least one radio relic is present. Their radio
morphology and location are quite varied, and could reflect different
physical origin or ICM conditions (Kempner et al. 2004; Giovannini \&
Feretti 2004). Due to low X-ray brightness at the cluster periphery, a
comparison of relic properties with the surrounding medium
(i. e. temperature and brightness gradient induced by shock waves) is
not obvious. Feretti \& Neumann (2006) did not find any evidence of a
temperature jump nearby the Coma cluster relic, and only recently a
temperature gradient has been found nearby the relic in Abell 548b by
Solovyeva et al. (2008); here a precise orientation of the cluster
merger with respect to the line of sight is inferred, and the
projected displacement of the relic from the shock is thus
explained.\\ Of particular interest is to explore the connection between
merger shock waves and clusters with double relics,
i.e. clusters hosting two relic radio sources located in the
peripheral region and symmetric with respect to the cluster center. So
far a very small number of clusters with two double relics has been
found. One of them is Abell 3667 (R\"ottgering et al. 1997;
Johnston-Hollitt et al. 2002). Here the cluster X-ray emission shows
an elongated shape, interpreted as the merger axis of two
sub-clusters, and relics are displaced symmetrically and perpendicular
to the main axis. X-ray, optical and radio properties have been
reproduced by a numerical simulation of a merger between clusters with
mass ratio of 0.2 by Roettiger et al. (1999). We note however that not
all of the predictions made by such simulations could be tested with
available data. Apart from Abell 3667, double relics have been
observed in Abell 3376 (Bagchi et al. 2006), and interpreted as
``Outgoing merger shock waves''. Double relics have also been observed
in RXCJ 1314.4-2515 (Feretti et al. 2005; Venturi et al. 2007), but no
detailed study on the relics formation has been performed on this
cluster so far. Two more candidates for hosting double relics are
Abell 2345 (Giovannini et al. 1999) and Abell 1240 (Kempner \& Sarazin
2001).\\
\begin{table*} [t!]
\caption{VLA observations}          
\label{tab:radioobs}      
\centering          
\begin{tabular}{|c c c c c c c c|}     % 8 columns 
\hline\hline       
Source       & RA           &  DEC      &$\nu$&Bandwidth&Config.& Date & Duration     \\
             & (J2000)      & (J2000)   &   (MHZ)  & (MHz)   &       &      & (Hours)  \\\hline                    
\hline
Abell 2345   & 21 27 12.0   & -12 10 30.0 &   325    & 3.125   &   B    &16-AUG-2006 & 2.0 \\% 147779\\ 
             &              &             &   325    & 3.125   &   C    &08-DEC-2006 & 5.4 \\%263514\\
Abell 1240   & 11 23 37.0   & 43 05 15.0  &   325    & 3.125   &   B    &05-AUG-2006 & 2.6\\% 214384\\
             &              &             &   325    & 3.125   &   C    &08-DEC-2006 & 4.7 \\%203675\\
\hline
Abell 2345-1 & 21 26 43.0  & -12 07 50.0 &    1425    & 50      &   C    &08-DEC-2006 & 1.9 \\%185755 \\%(24)
             & 21 26 43.0  & -12 07 50.0 &    1425    & 50      &   D    &09-APR-2007 & 1.0 \\%98376 \\%(24)
Abell 2345-2 & 21 27 36.0  &-12 11 25.0 &     1425    & 50      &   C    &08-DEC-2006 & 2.0 \\%196021\\%(24)
             & 21 27 36.0  &-12 11 25.0 &     1425    & 50      &   D    &09-APR-2007 & 1.0 \\%108368\\%(24)
Abell 1240-1 & 11 23 25.0  & 43 10 30.0 &     1425    & 50      &   C    &08-DEC-2006& 1.8 \\% 177879\\
             & 11 23 25.0  & 43 10 30.0 &     1425    & 50      &   D    &12-APR-2007& 1.0 \\%80364 \\
Abell 1240-2 & 11 23 50.0 & 43 00 20.0  &     1425    & 50      &   C    &08-DEC-2006& 1.9 \\%189945 \\
             & 11 23 50.0 & 43 00 20.0  &     1425    & 50      &   D    &12-APR-2007& 1.0 \\%78099\\                  
\hline
Abell 2345   & 21 26 57.2 & -12 12 49   &    1490     & 50      &  AnB   &02-NOV-1991& 0.1\\
\hline
\multicolumn{8}{l}{\scriptsize Col. 1: Source name; Col. 2, Col. 3: Pointing position (RA, DEC);
Col. 4: Observing frequency;}\\
\multicolumn{8}{l}{\scriptsize Col 5: Observing bandwidth; Col. 6: VLA configuration; 
Col. 7: Dates of observation; Col. 8: Net time on source.}\\ 
\end{tabular}
\end{table*}
We present here
new Very Large Array (VLA) observations of these two clusters at 20
and 90 cm to confirm and study the double relic emission in the
framework of relic formation models. Spectral index analysis of both
radio relics in the same cluster have not been performed so far. In
Abell 3667 the spectral index image has been obtained for only one of
the two relics, and no spectral index information are available for
relics in Abell 3376. Only integrated spectral index information are
available for the relics in RXCJ 1314.4-2515. Study of the spectral
index and of the polarization properties of relics offers a powerful
tool to investigate the connection between double relics and outgoing
shock waves originating in a merger event. In fact, theoretical models
and numerical simulations make clear predictions on the relic spectral
index trend and magnetic field properties (see Ensslin et al.
1998; Roettiger et al. 1999; Hoeft \& Br{\"u}ggen 2007).\\
The paper is
organized as follows: in Sec. 2 observations and data reduction are
described, in Sec. 3 and 4 we present the analysis of the cluster
Abell 2345 and Abell 1240.  Results are discussed in Sec. 5 in
comparison with other observed relics and theoretical models, and
conclusions are presented in Sec. 6. We assume a $\Lambda$CDM
cosmological model with $H_0=$71 km s$^{-1}$ Mpc$^{-1}$, $\Omega_M$=0.27,
$\Omega_{\Lambda}$=0.73.
%__________________________________________________________________

\section{VLA radio observations}
\subsection{Total intensity data reduction}
\label{sec:radioobs}
Observations have been performed at the Very Large Array (VLA) at 20
cm in the C and D configuration and at 90 cm in the B and C
configuration, in order to obtain the same spatial frequency coverage
in the UV plane. Observations details are given in
Tab. \ref{tab:radioobs}. \\ {\bf Observations at 20 cm (1.4 GHz)} have
been pointed separately on the two relics in both of the clusters
because of the smaller full width at half power of the primary
beam. Observations of the cluster Abell 1240 have been calibrated
using the source 3C286 as primary flux density calibrator\footnote{we
refer to the flux density scale by Baars \& Martin (1990)}. The source
1156+314 has been observed at intervals of about 30 min and used as
phase calibrator. Observations of Abell 2345 have been calibrated
using the sources 3C48 as primary flux density calibrator. Phase
calibration has been performed by observing the source 2137-207 at
intervals of $\sim$ 30 min.\\ We performed standard calibration and
imaging using the NRAO Astronomical Imaging Processing Systems
(AIPS). Cycles of phase self-calibration were performed to refine
antennas phase solutions, followed by a final amplitude and gain
self-calibration cycle.\\ In addition we recovered from the VLA data
archive a short observation performed with AnB array. The source 3C48
is used as primary flux density calibrator and the source 2121+053 is
used as phase calibrator. We reduced and calibrated these data as
explained above, details are given in
Tab. {\ref{tab:radioobs}}.  \\{ \bf Observations at 90 cm (325 MHz)}
have been performed in the spectral line mode, using 32 channels with
3.127 MHz bandwidth. This observing method avoids part of the VLA
internal electronics interferences and allows us to remove accurately
Radio Frequency Interferences (RFI). This also reduces bandwidth
smearing, that is quite strong at low frequencies. Primary flux
density and phase calibrators were the same sources used in 1.4 GHz
observations. 3C48 and 3C286 were also used for bandpass
calibration. RFI are particularly strong at low radio frequency, so
that an accurate editing has been done channel by channel, resulting
in a consistent flag of data. This in conjunction with bad data coming
from EVLA antennas results in a loss of $\sim$ 40 \% of observing
time.  Calibration has been performed following the ``Suggestions for
P band data reduction'' by Owen et al. (2004).\\ After the initial
bandpass calibration channels from 1 to 4 and from 28 to 32 have been
flagged because of the roll-off of the bandpass. In the imaging
procedure data have been averaged to 8 channels. Imaging has been
performed using the wide field imaging technique to correct for non
complanarity effects over a wide field of view. 25 facets covering
the main lobe of the primary beam have been used in the cleaning and
phase-self calibration processes. We also searched in
the NVSS data archive for sources stronger then 0.5 Jy over a radius
as large as 10$^{\circ}$. These sources have been included in the
initial cleaning and self calibration steps.\\ Each (u,v) data set at
the same frequency but observed with different configurations has been
calibrated, reduced and imaged separately and then combined to produce
the final images. Images resulting from the separate pointed
observations at 1.4 GHz have been then linearly combined with the AIPS
task LTESS. We combined the data set and produced images at higher and
lower resolution (herein after HR images and LR images) giving uniform
and natural weight to the data. For the purposes of the spectral
analysis, the final images at 325 MHz and 1.4 GHz, have been restored
with the same beam (reported in Tab. \ref{tab:2345radiofinali} and
\ref{tab:1240radiofinali}) and corrected for the primary beam effects.
\begin{table}[h!]
\caption{ Abell 2345}
\label{tab:2345radiofinali}
\centering
\begin{tabular} {|l c c c c|} 
\hline\hline
Source name & $\nu$ & $\theta$   & $\sigma_{I}$ & $~~~$Fig. \\
             &   MHz     & arcsec & mJy/beam  & \\
\hline
Abell 2345-1 HR &  1425      & 37 X 20    &  0.08 & \\
Abell 2345-1 LR &  1425      & 50 X 38    &  0.09 & \ref{fig:A2345_spixcut}, central panel\\
Abell 2345-2 HR &  1425      & 37 X 20    &  0.09 & \\
Abell 2345-2 LR &  1425      & 50 X 38    &  0.09 & \ref{fig:A2345_spixcut}, central panel\\
Abell 2345   HR &  325      & 37 X 20    &   1.7  & \\
Abell 2345   LR &  325      & 50 X 38    &   2.0  & \ref{fig:A2345_spixcut}, right panels\\
\hline
Abell 2345    &  1490     &6X6         &   0.13  & \ref{fig:A2345_otticoradio}, central panel\\
\hline
\multicolumn{5}{l}{\scriptsize Col. 1: Source name; Col. 2: Observation frequency;
}\\
\multicolumn{5}{l}{\scriptsize Col. 3: Restoring beam; Col. 4: RMS noise of the 
final images;}\\
\multicolumn{5}{l}{\scriptsize  Col 5: Figure of merit.}
\end{tabular}
\end{table}
\begin{table}[h!]
\caption{ Abell 1240}
\label{tab:1240radiofinali}
\centering
\begin{tabular} {|l c c c c|} 
\hline\hline
Source name & $\nu$ & $\theta$   & $\sigma_{I}$ & $~~~$Fig. \\ 
             &   MHz     & arcsec & mJy/beam &\\
\hline
Abell 1240-1 HR &  1425      & 22 X 18    &  0.04&\ref{fig:A1240_otticoradio.ps} \\
Abell 1240-1 LR &  1425      & 42 X 33    &  0.04&\ref{fig:A1240spix},central panel \\
Abell 1240-2 HR &  1425      & 22 X 18    &  0.04&\ref{fig:A1240_otticoradio.ps} \\
Abell 1240-2 LR &  1425      & 42 X 33    &  0.05&\ref{fig:A1240spix}, central panel \\
Abell 1240   HR &  325      & 22 X 18    &  0.9 &\\
Abell 1240   LR &  325      & 42 X 33    &  1.0& \ref{fig:A1240spix}, left panels\\
\hline
\multicolumn{5}{l}{\scriptsize Col. 1: Source name; Col. 2: Observation frequency;}\\
\multicolumn{5}{l}{\scriptsize Col. 3: Restoring beam; 
                    Col. 4: RMS noise of the final images;}\\
\multicolumn{5}{l}{\scriptsize Col. 5: Fig. of merit.}
\end{tabular}
\end{table}
\begin{table}[h!]
\caption{Total and polarization intensity radio images at 1425 MHz}
\label{tab:radiopol}
\centering
\begin{tabular}{|c  c c c  c| } 
\hline\hline 
Source name & $\theta$ & $\sigma_I$ & $\sigma_{Q,U}$ & Fig. \\ 
           & arcsec & \scriptsize(mJy/beam) & \scriptsize(mJy/beam)&\\ 
\hline 
Abell 2345-1 & 23 X 16 & 0.05 & 0.02 & \ref{fig:A2345_pol},
right panel\\ Abell 2345-2 & 23 X 16 & 0.07 & 0.02 &
\ref{fig:A2345_pol}, left panel\\ Abell 1240-1 & 18 X 17 & 0.04 & 0.02
&\ref{fig:A1240pol}, top panel\\ Abell 1240-2 & 18 X 17 & 0.04 & 0.01
&\ref{fig:A1240pol}, bottom panel \\ \hline
\multicolumn{5}{l}{\scriptsize Col. 1: Source name; Col. 2: Restoring
beam; }\\ \multicolumn{5}{l}{ \scriptsize Col. 4: RMS noise of the I
image; Col 5: RMS noise of the Q and U images}\\
\multicolumn{5}{l}{\scriptsize Col 6: Figure of merit.}
\end{tabular}
\end{table}
\subsection{Polarization intensity data reduction}
Observations at 20 cm (1.425 GHz) include full polarization
information. Polarization data observed with the D array are unusable
because of bad quality of data of the polarization calibrator.  The
absolute polarization position angle has been calibrated by observing
3C286 for both clusters in C configuration. The instrumental
polarization of the antennas has been corrected using the source
1156+314 for Abell 1240 and the source 2137-207 for Abell 2345.\\
Stokes parameters U and Q images have been obtained. We then derived
the Polarization intensity image ($P=\sqrt{U^2+Q^2}$), the
Polarization angle image ($\Psi=\frac{1}{2}arctan\frac{U}{Q}$) and the
Fractional Polarization image ($FPOL=\frac{P}{I}$), with I being the
total intensity image. Further details are given in
Tab. \ref{tab:radiopol}.\\
\begin{table*}
\caption{Abell 2345 and Abell 1240 properties}          
\label{tab:x}      
\centering          
\begin{tabular}{|c c c c c c c |}    
\hline\hline       
Source name &  RA          &   DEC       & z     &  scale    & F$_X$                   & L$_X$ \\
            & (J2000)      & (J2000)     &       & (kpc/$''$) & $10^{-12}$ erg/s/cm$^2$ & $10^{44}$ erg/s \\
\hline
Abell 2345  &  21 27 11.00 & -12 09 33.0 & 0.1765& 2.957     &  5.3                & 4.3\\
Abell 1240  &  11 23  32.10 & 43 06 32   & 0.1590& 2.715     &  1.3               &  1.0\\
\hline
\multicolumn{7}{l}{\scriptsize Col. 1: Source name; Col. 2, Col. 3: Cluster X-ray centre (RA,
DEC); Col 4: Cluster redshift; Col 5: arcsec to kpc conversion scale;} \\
\multicolumn{7}{l}{\scriptsize Col 6: Flux in the 0.1-
2.4 keV band (Abell 2345) and in the 0.5-2 keV (Abell 1249); Col 7: X-ray cluster luminosity   }\\
\multicolumn{7}{l}{\scriptsize in the 0.1-2.4 keV band (Abell 2345) and in the 0.5-2 keV (Abell 1240);}\\
\multicolumn{7}{l}{\scriptsize Data from B\"ohringer et al. (2004) for Abell 2345 and from
David et al. (1999) for  Abell 1240, corrected for the adopted cosmology.}
\end{tabular}
\end{table*}
\begin{figure*}[t!]
\centering
\includegraphics[width=19.5cm]{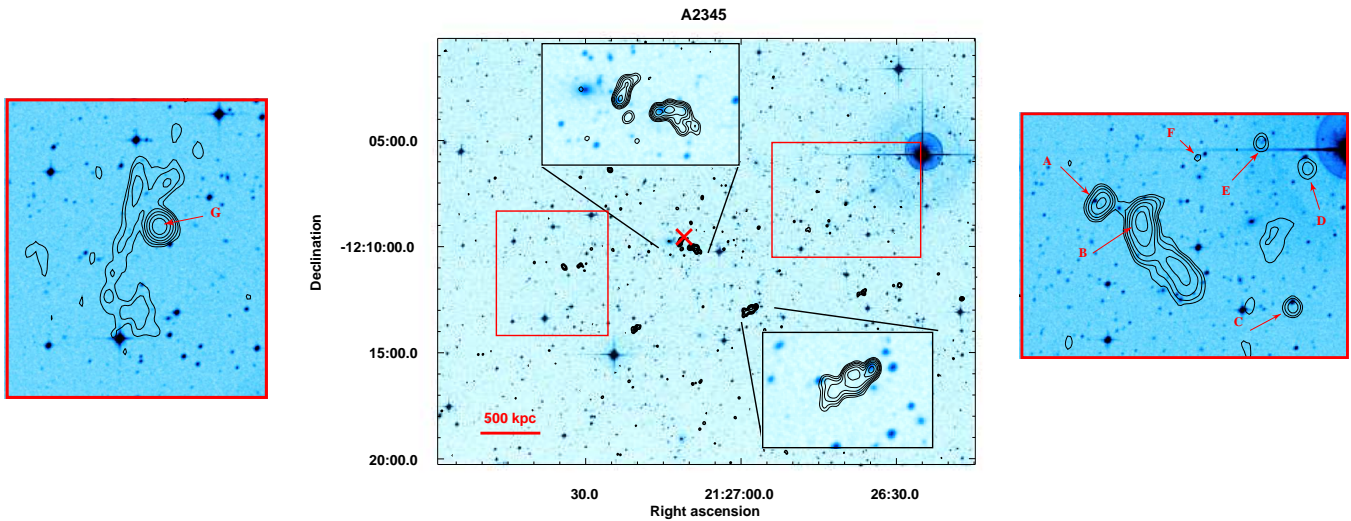}
\caption{The cluster Abell 2345. In the center: DPOSSII optical
emission (red band) in colors overlaid onto radio contours at 1.490
GHz. First contours are $\pm$ 0.4 mJy/beam and are then spaced by a
factor 2. The beam in 6$''$$\times$6$''$. Top inset shows the zoomed
images of the central sources: the central cD galaxy and two radio
galaxies are visible. Bottom inset shows the zoomed image of the
southern radio source. Red Boxes mark the region of the relics,
completely resolved in the high resolution image. In the Left and
Right panels zoomed image of the red boxes is shown. Here colors
represent the optical DPOSSII emission, while contours represents the
relic radio emission at the resolution of 23$''\times$16$''$. The
relic A2345-1 is visible in the right panel, while A2345-2 is in the
left panel. Contours start at $\pm$0.15mJy/beam and are spaced by a
factor 2. Red arrows indicate the position of the discrete sources
embedded in the relic emission.}
\label{fig:A2345_otticoradio}
\end{figure*}
\section{The Cluster Abell 2345}
\begin{figure*}[t!]
\centering
\includegraphics[width=19cm]{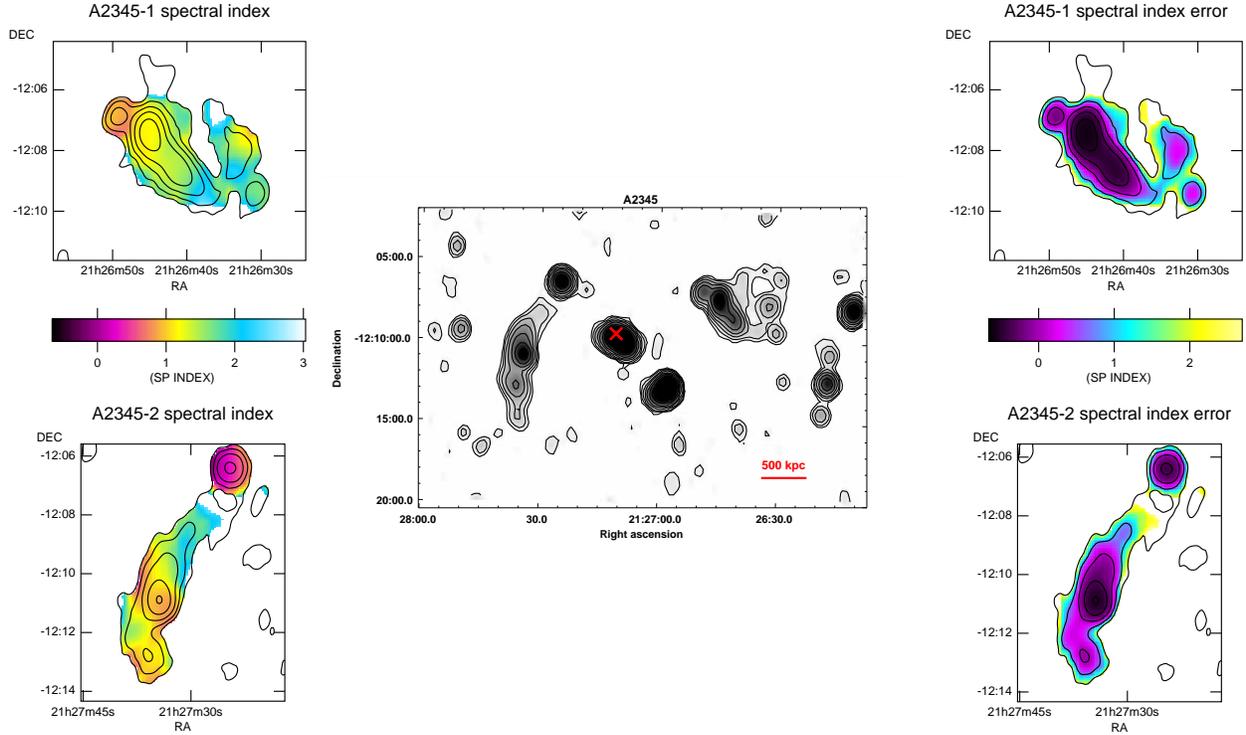}
\caption{Center: the cluster Abell 2345 radio emission at 1.4 GHz. The
beam is 50$''$$\times$38$''$. Contours start at 3 $\sigma$ (0.24
mJy/beam) and are then spaced by a factor 2. The cross marks the X-ray
cluster center. Left: Colors represent the spectral index of the
relic A2345-1 (top) and A2345-2 (bottom) superimposed over the radio
emission at 325 MHz (contours). The beam is 50$''$$\times$38$''$,
contours start at 3$\sigma$ (6 mJy/beam) and are then spaced by a
factor 2. Right: Spectral index error image (colors) superimposed onto
the emission at 325 MHz (contours are as above). }
\label{fig:A2345_spixcut}
\end{figure*}

Optical information are available for this cluster, while little is
known about its X-ray emission. General data are reported in
Tab. \ref{tab:x}.\\
Weak gravitational lensing analysis has been performed by Dahle et al.
(2002) and by Cypriano et al. (2004). Optical data cover the inner
part of the cluster ($\sim$3$'$$\times$3$'$). They find that this cluster
has a well defined core dominated by a cD galaxy, and both the light
and galaxy number density distribution have several peaks close to the
central galaxy.  The authors derived that the projected mass
distribution has the most prominent peak displaced from the central cD
by $\sim$1.5$'$, although a secondary peak is closer to the central
cD. No information about the possible presence of a cooling flow
associated with this galaxy is present in the literature. Dahle et
al. (2002) conclude from their analysis that the cluster may be a
dynamically young system. Cypriano et al. (2004) report the mass
distribution derived from weak lensing analysis and find that the best
fit to their data is a singular isothermal ellipsoid with the main
axis oriented in the E-W direction.\\
\smallskip\\ The radio emission of Abell 2345 is characterized by the
presence of two relics visible in the NVSS (Giovannini et al. 1999).\\
Our new VLA observations confirm the presence of two regions where
non-thermal emission is present at the cluster periphery, nearly
symmetrical with respect to the cluster center.  These new
observations together with the archive data, allow the study of the
cluster radio emission in a wide range of resolutions going from
$\sim$6$''$ to $\sim$50$''$. Therefore, it is possible to separate the
contribution of discrete sources whose emission is not related to the
relic's physical properties. In Fig. \ref{fig:A2345_otticoradio} the
radio emission of Abell 2345 at 6$''$ resolution is shown overlaid
onto the optical emission (taken from the Digitalized Palomar Sky
Survey II, red band). Two central radio-tail sources are associated
with optical galaxies in the cluster center. The central cD is visible
in the optical image. Relics are not visible in this image because of
the lack of short baselines. This confirms that the emission detected
in lower resolution observations is indeed extended and it is not due
to the blending of discrete sources. In the same figure we also report
the radio relic emission as detected by C array observations.  The
western relic (Abell 2345-1) is located at $\sim$ 1 Mpc from the
cluster X-ray center while the eastern relic (Abell 2345-2) is $\sim$
890 kpc far from the cluster center (see
Tab. \ref{tab:A2345_relics}).\\ There are several discrete sources in
proximity of the western relic, A2345-1, visible in the 1.4 GHz image,
they are labeled with letters from A to F in the right panel of
Fig.\ref{fig:A2345_otticoradio}. The sources A, C, D, E and F could be
associated with the optical galaxies visible in the DPOSSII image,
whereas B does not have any obvious optical identification.  Optical
emission is present at 35$''$ in NE direction from the radio
peak. This is larger then the error associated with the beam, that is
only 6$''\times$6$''$ in the highest resolution image. We can then
conclude that no optical counterpart of the B radio source is detected
in the DPOSSII image. The sources D E and F are not visible in the 325
MHz image (see Fig. \ref{fig:A2345_spixcut}, top left panel). This is
consistent with a radio source having a spectral index
$<1.2$\footnote{The spectral index $\alpha$ is derived according to
$S_{\nu} \propto \nu^{-\alpha}$.}. There is only one discrete source
in proximity of the relic A2345-2, labeled with G in the
Fig.\ref{fig:A2345_otticoradio} without any obvious optical
identification. This source is also detected in the higher resolution
image. \\ The whole extension of the relics is properly revealed by LR
images (Fig. \ref{fig:A2345_spixcut}).  The morphology of the relics
is similar at 1.4 GHz and 325 MHz, although only the brightest regions
can be seen at 325 MHz due to the higher rms noise level of these
observations with respect to the 1.4 GHz ones.  The total flux of the
relics at the 2 frequencies, excluding the contribution of the
discrete sources, are reported in Tab. \ref{tab:A2345_relics}, where
the main physical parameters are summarized.\\ The relic A2345-1 shows
an elongated shape at high resolution, while at lower resolution it
shows a weaker wide emission extending in the West direction
i.e. toward the cluster outskirts. We note that this circular
filamentary morphology is not seen in other double relic sources, as
discussed in the introduction.\\
\begin{figure*}[t!]
   \label{fig:A23451spixprofile.ps}
   \centering
   \includegraphics[width=9cm]{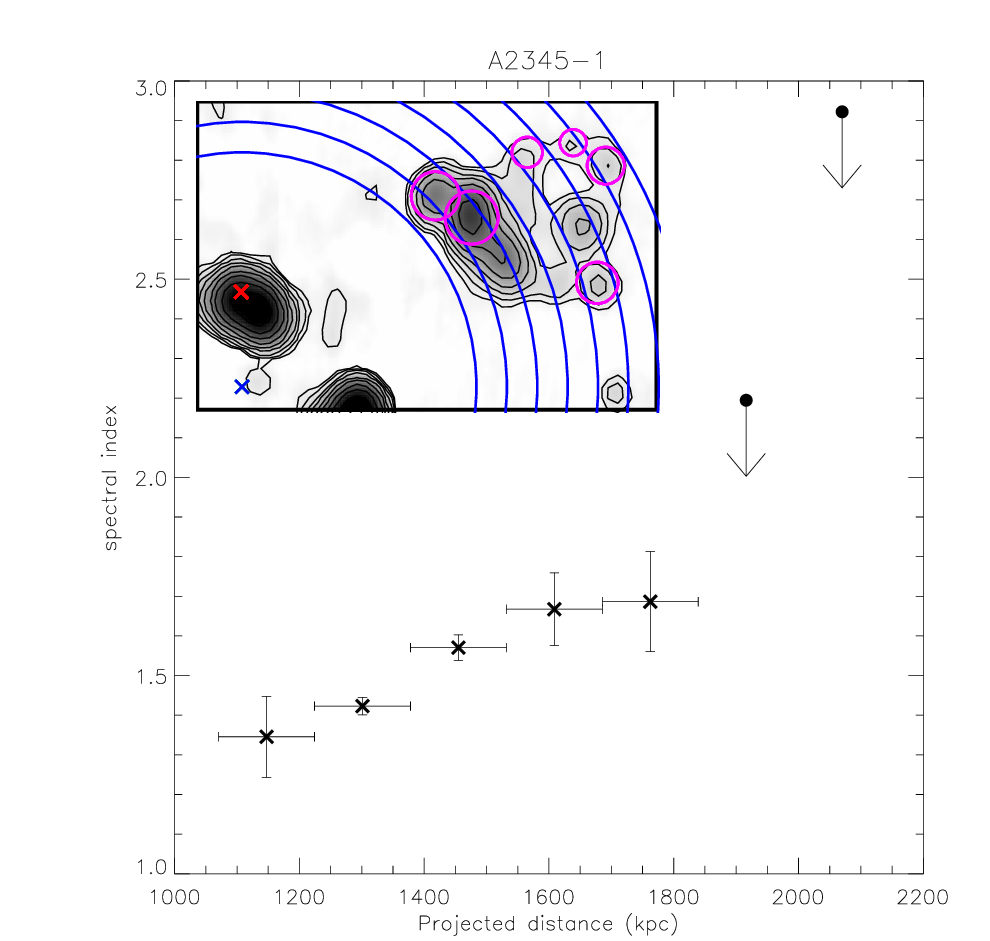}
    \includegraphics[width=9cm]{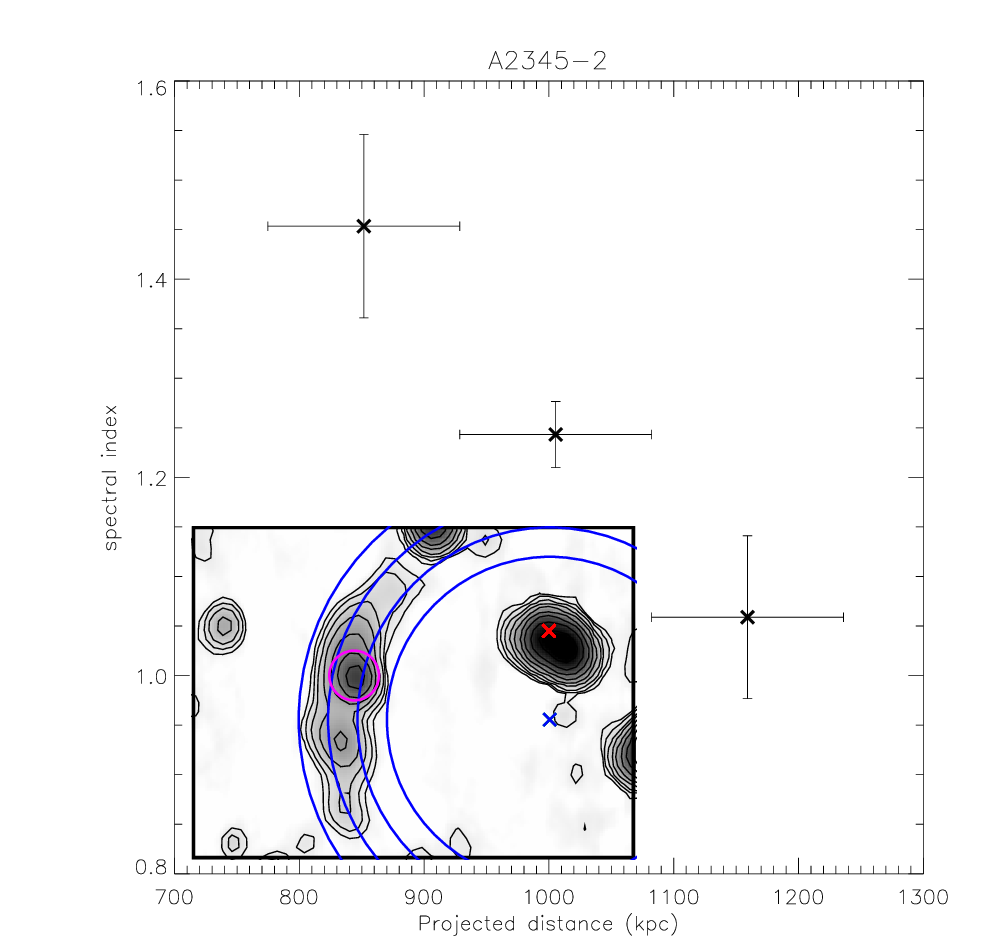}
  \caption{Spectral index radial trend of A2345-1 (left) and A2345-2
          (right), computed in shells of $\sim 50''$ in width. It has
          been computed excluding the contribution of the discrete
          sources. Crosses refer to spectral index values computed in
          shells where the mean brightness is $>3\sigma$ at both 325
          MHz and 1.4 GHz. Arrows are $3\sigma$ upper limits on the
          spectral index mean value (see text). The red cross refers
          to the cluster X-ray center, the blue cross refers to the
          center of the spherical shells. In the insets: displacement
          of the shells over which the mean spectral index has been
          computed. Circles refer to the discrete sources embedded in
          the relic emission. The red cross refers to the cluster
          X-ray center, the blue cross is the center of the spherical
          shells. }
   \end{figure*}
\subsection{Spectral index analysis}
\label{A2345spectralindex}
We derived the spectral index image of the cluster's relics comparing
the LR images at 1.4 GHz and 325 MHz. The rms noise of the images are
reported in Tab. \ref{tab:2345radiofinali}. Spectral index and
spectral index noise images are shown in
Fig. \ref{fig:A2345_spixcut}. They have been obtained by considering
only pixels whose brightness is $> 3\sigma$ at both frequencies. We
note that relics are more extended at 1.4 GHz than at 325 MHz. This
can be due to the different sensitivities at 1.4 GHz and 325
MHz. Confusion and RFI strongly affect the low frequency image, where
the noise level is significantly higher than the thermal noise. A
consistent spectral index analysis has to consider the different
extension at the two frequencies. In fact, as already pointed out by
Orr\'u et al. (2007) if we compute spectral index analysis considering
only regions that have a signal to noise ratio $>$ 3 at both
frequencies, we introduce a bias, since we are excluding a priori low
spectral index regions, whose emission cannot be detected at 325
MHz. For instance, the relic A2345-1 radio brightness at 1.4 GHz
decreases as the distance from the cluster center increases. The
fainter region could be detected in the 325 MHz image only if its
spectral index, $\alpha$, were steeper than $\sim$1.8.\\ In both of
the relics the spectral index is patchy. The spectral index rms is
$\sigma_{spix}\sim$ 0.4 while the mean spectral index noise is $<$Spix
Noise$>\sim$ 0.1 for both relics. Thus, by comparing these two
quantities we can conclude that spectral index features are
statistically significant.  \\ Our aim here is to investigate if there
is a systematic variation of the relic spectral index with distance
from the cluster center as found in other radio relics (e.g. 1253+275
by Giovannini et al. 1991; Abell 3667 by R\"ottgering et al. 1997;
Abell 2744 by Orr\'u et al. 2007; Abell 2255 by Pizzo et al. 2008;
Abell 521 by Giacintucci et al. 2008; ).\\ In order to properly
obtain the radial trend of the spectral index, we integrated the radio
brightness at 325 MHz and 1.4 GHz in radial shells of $\sim 50''$ in
width wherever the 1.4 GHz brightness is $>3\sigma$, and then we
computed the value of the spectral index in each shell. We have
excluded the regions where discrete radio sources are embedded in the
relic emission (see insets in Fig. \ref{fig:A23451spixprofile.ps}).
The shells have been centered in the extrapolated curvature center of
the relic A2345-2, that is 2.6$'$ South the cluster X-ray
center. Shells are then parallel to the relics main axis. We computed
the integrated brightness in each shell at 20 and 90 cm , and
calculated the associated error as $\sigma\times\sqrt{N_{beam}}$,
where $\sigma$ is the image rms noise, $N_{beam}$ is the number of
beams sampled in the shell.  In those shells where the brightness is
$>3\sigma$ in the 1.4 GHz image but $<3\sigma$ in the 325 MHz image,
only upper limits on the mean spectral index can be derived. The
spectral index profiles thus obtained are shown in
Fig. \ref{fig:A23451spixprofile.ps}. These plots shows that the
spectral index in the relic A2345-1 increases with distance from the
cluster center, indicating a spectral steepening of the emitting
particles. The spectral index, in each shell, is rather high, going
from $\sim$1.4 in the inner rim to $\sim$1.7 in the central rim of the
relic. The spectral index trend derived for the outer shells is
consistent with a further steepening.\\ The spectral index of the
relic A2345-2 shows instead a different trend, going from $\sim$1.4 in
the inner shell to $\sim$1.1 in the outer rim
(Fig. \ref{fig:A23451spixprofile.ps}).
\begin{figure*}[t!]
\centering
\includegraphics[width=1.2\columnwidth]{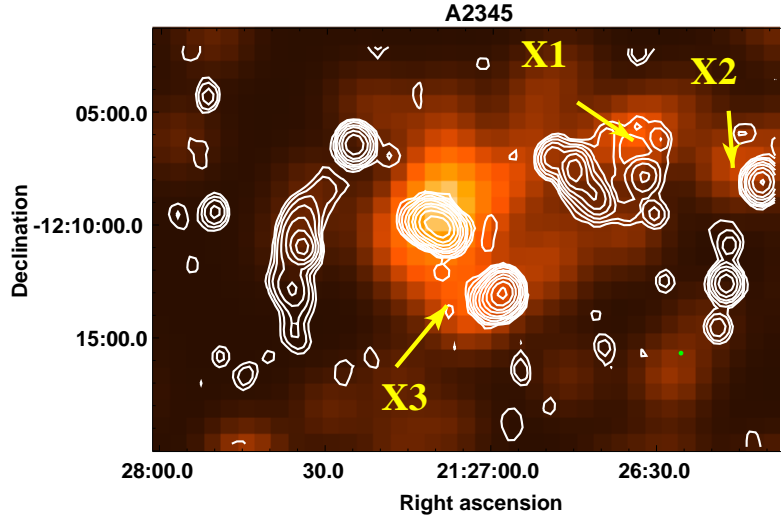}
\caption{Abell 2345 X-ray emission (colors) in the energy band 0.1-2.4
keV from ROSAT PSPC observations. The image has been smoothed with a
Gaussian of $\sigma\sim$60$''$;
contours represent the radio image of the cluster at 1.4 GHz. The beam
is 50$''$$\times$38$''$. Contours are 0.24 mJy/beam and are then spaced
by a factor 2. Arrows mark the position of the X1 X2 and X3 regions. }
\label{fig:A2345X}
\end{figure*}
\subsection{Radio-X-ray comparison}
No X-ray studies are present in the literature for this cluster.
X-ray observations in the energy band $0.1-2.4$ keV have been
retrieved by the ROSAT data archive. The cluster is $\sim$ 100$'$
offset from the ROSAT pointing. Observations have been performed with
the ROSAT PSPC detector for a total exposure time of $\sim$ 14
ksec. After background subtraction the event file has been divided by
the exposure map. We smoothed the resulting image with a Gaussian of
$\sigma=60''$. The resulting image is shown in
Fig. \ref{fig:A2345X}.\\ The X-ray emission of this cluster is
elongated in the NW-SE direction. Two bright regions are visible at
$\sim$ 10$'$ and 14$'$ in N-W direction from the cluster center (we
refer to them as X1 and X2 respectively). The galaxy J21263466-1207214
(RA =21h26m34.6s, DEC= -12d07m22s, z=0.178221) is close to the first
one. Another bright region is present at $\sim$ 4$'$ South from the
cluster center (X3). \\ Data presented here allow an interesting
comparison among cluster emission at different wavelengths. We note
that mass distribution from weak lensing studies (Cypriano et
al. 2004) is well represented by an ellipsoid with the major axis
directed in the E-W direction and relics are found perpendicular to
this axis. Consistently with the optical analysis, the X-ray emission
is elongated in the NW-SE direction, indicating a possible merger
along that direction, and relics are displaced perpendicular to that
axis. In Fig. \ref{fig:A2345X} the X-ray emission is superimposed onto
radio contours. A2345-2 is located at the edge of the X-ray emission,
as found in relics of Abell 3667 and A3376. A2345-1, instead, is
located between eastern edge of the cluster and the X1 region, 10$'$
from Abell 2345 center, and its radio emission extends toward X1.\\
From the same figure, in the X3 region a Narrow Angle Tail radio
galaxy is visible in radio images at every resolution (see
Fig.\ref{fig:A2345_otticoradio} and \ref{fig:A2345_spixcut}). Although
redshift is not available for this radio source, its structure favors
a connection to the cluster and/or to the close X-ray peak. One
possibility is that these X-ray multiple features are galaxy clumps
interacting with Abell 2345. \\ A self consistent scenario arises from
this analysis, indicating that the cluster Abell 2345 could be
undergoing multiple merger with X3 and X1 groups, and this could
explain the peculiar properties of A2345-1. More sensitive and
resolved X-ray observations in conjunction with optical studies are
required to shed light on the connection between the radio emission of
A2345, X1 X2 and X3.\\
\subsection{Equipartition Magnetic field}
\label{secs:Equip}
Under the assumption that a radio source is in a minimum energy
conditions, it is possible to derive an average estimate of the
magnetic field strength in the emitting volume (see e.g. Pacholczyk
1970). We assume that the magnetic field and relativistic particles
fill the whole volume of the relics, and that energy content in
protons and electrons is equal. We further assume that the volume of
the relics is well represented by an ellipsoid having the major and
minor axis equal to the largest and smallest linear scale visible in
our images; we estimated the third axis to be the mean between the
major and minor one. The synchrotron luminosity is calculated from a
low frequency cut-off of 10 MHz to a high frequency cut-off of 10
GHz. The emitting particle energy distribution is assumed to be a
power law in this frequency range ($N(E)\propto E^{-p}$), with
$p=2\alpha+1$. We used the mean value of $\alpha=$1.5 and 1.3 for
A2345-1 and A2345-2 respectively and found $B_{eq}\sim$1.0 $\mu$G in
A2345-1 and 0.8 $\mu$G in A2345-2. These values are consistent with
equipartition magnetic field found in other relics.\\ It has been
pointed out by Brunetti et al. (1997) that synchrotron luminosity
should be calculated in a fixed range of electron energies rather than
in a fixed range of radio frequencies (see also Beck \& Krause
2005). In fact, electron energy corresponding to a fixed frequency
depends on the magnetic field value, and thus the integration limits
are variable in terms of the energy of the radiating particles. Given
the power law of the radiating particles and the high value of the
radio spectral index, the lower limit is particularly relevant
here. We adopted a low-energy cut off of $\gamma_{min}$=100 and
assumed $\gamma_{max}>>\gamma_{min}$, obtaining $B'_{eq}\sim$ 2.9
$\mu$G in A2345-1 and 2.2 $\mu$G in A2345-2.\\ We derived the minimum
non thermal energy density in the relic sources from $B'_{eq}$
obtaining $U_{min}\sim$8.1 and 4.3 10$^{-13}$erg/cm$^{-3}$ for A2345-1
and A2345-2. The corresponding minimum non-thermal pressure is then
$\sim$5.0 and 2.7 10$^{-13}$erg/cm$^{-3}$. \\ We are aware that
  the extrapolation to low energies or frequencies could over estimate
  the number of low energy electrons, leading to over estimate the
  equipartition magnetic field if a spectral curvature is present.  We
  note that a detailed study of the radio spectrum on a large
  frequency range is available for three peripheral relics: the one in
  Abell 786, in the Coma cluster (see Giovannini \& Feretti, 2004 and
  references therein) and in Abell 521 (Giacintucci et al. 2008). In
  these relics a straight steep radio spectrum is observed. We also note that a
  low frequency cut-off of 10 MHz and a magnetic field of
  $\sim$1$\mu$G imply a low energy cut-off of
  $\gamma_{min}\sim$1500. Thus, also if the spectrum of the emitting
  particles is truncated at $\gamma>$1500, both $B'_{eq}$ and $B_{eq}$
  could over estimate the magnetic field strength. Future
  low-frequency radio interferometers such as SKA and LWA will likely shed
  light on this point. On the other hand, it is possible to derive an
  independent estimate of the magnetic field from X-ray flux due to
  inverse Compton scattering of CMB photons by relativistic electrons
  in the relic source. These studies have been performed so far on a
  scarce number of radio relics and have led to lower limits on the
  magnetic field strength: $B>$0.8$\mu$G in the relic 1140+203 of
  Abell 1367 (Henriksen \& Mushotzky 2001); $B>$1.05$\mu$G in 1253+275
  of the Coma cluster (Feretti \& Neumann 2006; $B>$0.8$\mu$G in
  0917+75 in Rood27 cluster (Chen et al. 2008); and $B>$2.2$\mu$G in
  the relic 1401-33 in the Abell S753 cluster (Chen et al. 2008). In
  these cases, the lower limits derived from IC arguments are
  consistent with equipartition estimates, thus indicating that the
  equipartition value could be used as a reasonable approximations of
  the magnetic field strength in relics.
\begin{table*} [t]
\caption{ Abell 2345}
\label{tab:A2345_relics}
\centering
\begin{tabular} {|c  c c c c c c  |} 
\hline\hline
Source name &  Proj. dist & LLS         & F$_{20 cm}$   &  F$_{90 cm}$ & B$_{eq}$ - B$'_{eq}$& $<\alpha>$ \\ 
            &  kpc        & kpc         & mJy          &  mJy        & $\mu$G     &             \\
\hline
Abell 2345-1 &  340$''$=1000& 390$''$= 1150 &30.0$\pm$0.5& 291$\pm$ 4 &  1.0 -2.9  &    $1.5\pm$0.1  \\
Abell 2345-2 &  300$''$=890 & 510$''$= 1500 &29.0$\pm$0.4& 188$\pm$ 3 &  0.8 -2.2   &    $1.3\pm$0.1  \\

\hline \multicolumn{7}{l}{\scriptsize Col. 1: Source name; Col. 2:
projected distance from the X-ray centroid; Col. 3: Largest linear
scale measured on the 20 cm images.}\\ 
\multicolumn{7}{l}{\scriptsize Col. 4 and 5: Flux density at 20 and 90 cm; Col. 6: equipartition
magnetic field computed at fixed frequency - fixed energy }\\
 \multicolumn{7}{l}{\scriptsize (see Sec. \ref{secs:Equip}); Col. 7: mean spectral index in
region where both 20 and 90 cm surface brightness is $>$ 3 $\sigma$}
\end{tabular}
\end{table*}
\begin{figure*}[t!]
   \centering \includegraphics[width=18cm]{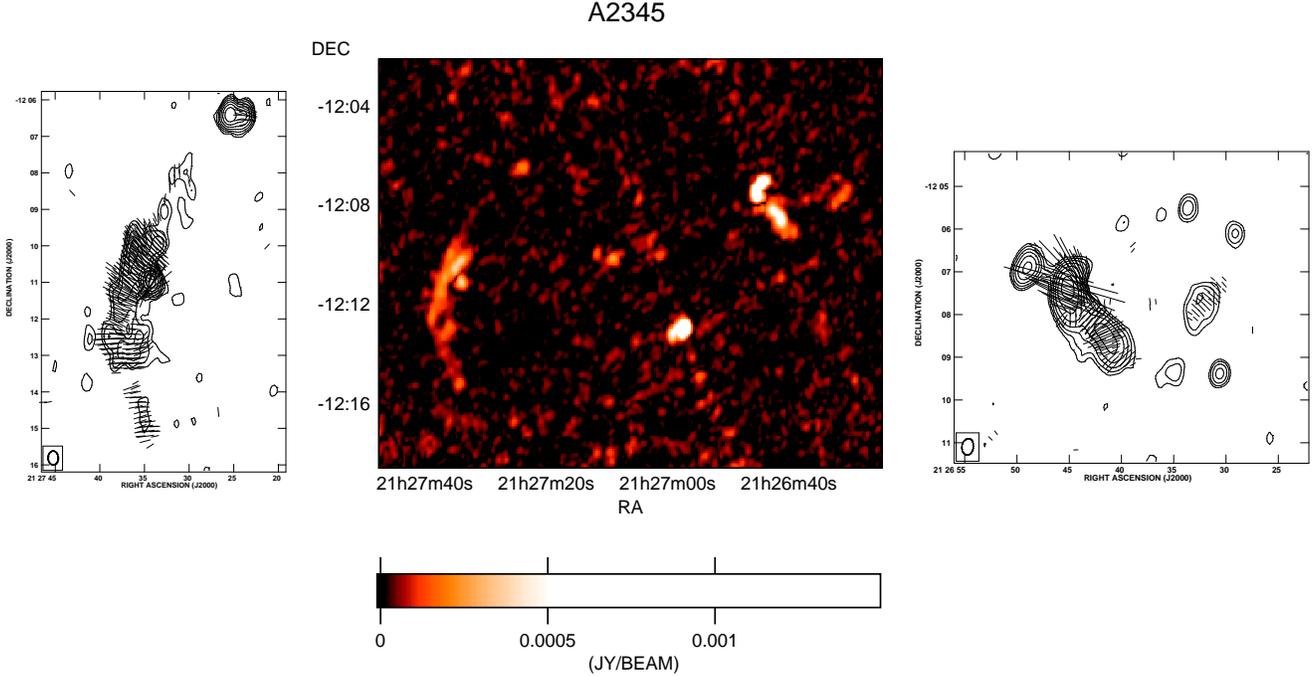}
      \caption{Abell 2345: Polarized emission of Abell 2345 at 1,4
      GHz. In the central panel the polarized radio emission at 1.4 GHz
      is shown. The restoring beam is $23''\times 16''$. Left and
      right panels: contours refer to the radio image Abell 2345-2 and
      Abell 2345-1 (see Tab. \ref{tab:radiopol} for further
      details). Contours start from $3 \sigma$ and are spaced by a
      factor 2. E vectors are superimposed: line orientation indicates
      the direction of the E field, while line length is proportional
      to the polarization intensity (Left panel: 1$''$ corresponds to
      5.5 $\mu$Jy/beam; Right panel: 1$''$ corresponds to 10
      $\mu$Jy/beam)}
         \label{fig:A2345_pol}
   \end{figure*} 
\subsection{Polarization analysis}
\label{sec:A2345pol}
Another important set of information about the magnetic field in the
relics can be derived through the study of polarized emission. As
previously mentioned, we could calibrate polarization only for
observations at 1.4 GHz with the C array.\\ In Fig. \ref{fig:A2345_pol} the P radio image of
the cluster is shown. The noise achieved in the P, Q and U images
(Tab. \ref{tab:radiopol}) are lower than those obtained in total
intensity image. In fact total intensity image are affected by
dynamical range limitation due to the presence of powerful radio
sources near our target. These sources are not strongly polarized, so
that P images are not affected by such limitation, and weaker
polarized emission can be revealed. We note in fact that polarized
radio emission of the relic A2345-2 reveals an arc-like structure that
is more extended than in total intensity emission. The arc-like
structure of this relic indicates that the shock wave, possibly
responsible of the radio emission, has been originated $\sim 2.6'$
southern the present X-ray center.\\ The mean fractional polarization
is $\sim 22 \%$ in A2345-2, reaching values up to 50\% in the eastern
region. The relic A2345-1, shows a mean fractional polarization of
$\sim 14 \%$ with higher polarized region ( $\sim 60 $\%) in the
north-western part of the relic.  The amount of fractional
polarization allows to estimate the level of order of the magnetic
field in the source. Following Burn (1966), if we assume that the
magnetic field is composed by an ordered component ${\bf B_o}$ plus a
random isotropic component represented by a Gaussian with variance
equal to $2/3 B_{r}^2$, it results
\begin{equation}
P_{oss}=P_{int}\frac{1}{1+(B_r^2/B_o^2)}
\label{eq:pol}
\end{equation}
 where $P_{oss}$ is the observed fractional polarization, while
$P_{intr}$ is given by $P_{intr}=\frac{3\delta+3}{3\delta+7}$.  For
the relic A2345-1 we obtain that $B_r^2/B_o^2 \sim 4$, meaning that
the magnetic energy density in the random component is three times
larger than the one in the ordered component. For the relic A2345-2,
instead, we obtain that $B_r^2/B_o^2 \sim 2$. This indicates that the
magnetic field in the region of the relic A2345-2 has a higher degree
of order. We also have to consider possible beam depolarization,
internal depolarization and ICM depolarization, so that what we can
conclude from this analysis is $B_r^2/B_o^2 < 4$ and $<2$ in A2345-1
and A2345-2 respectively.\\ In A2345-1 the magnetic field is
mainly aligned with the sharp edge of the radio emission, i.e. in the
SW-NE direction. In the northern part of the relic the E vectors rotate
and in the N-W part they are almost aligned toward the SW-NE
direction. In A2345-2 the E vectors are perpendicular to the relic
major axis, following the arc-like structure that is marginally
visible in the total intensity image. \\
\subsection{Results on Abell 2345}
\label{Sec:A2345results}
The presented analysis confirms that non-thermal emission is
associated with the ICM of Abell 2345.
\begin{itemize}
\item {The properties of the western relic, A2345-1 are quite
  peculiar.   We note indeed that its morphology is rather
    circular and filamentary, its brightness distribution is higher in
    the inner region of the relic and its spectral index steepens
    toward the cluster periphery. Although the statistic is really
    poor, these features have not been found in other double relics so
    far. The level of polarization, the magnetic field direction
    mainly aligned with the sharp edge of the radio emission, and the
    value of the equipartition magnetic field are instead in agreement
    with other observed relics.\\ Diffusive shock acceleration models
    predict a steepening of the radio spectrum towards the cluster
    center (e.g. Ensslin et al. 1998; Hoeft \& Br{\"u}ggen 2007) as a
    consequence of the electron energy losses after shock
    acceleration. It is worth mentioning here that theoretical
    predictions rely on some assumptions about the
    shock symmetry and the magnetic field structure that could be not
    representative of this specific cluster environment.  Moreover, if
    the relic is not seen edge on, projection effects could further
    complicate the observed radio emission. Taking all of these into
    account, the observed spectral index trend of A2345-1 cannot be
    used as an argument to exclude an outgoing shock wave. \\ We
    however note that the position of A2345-1 is in between the main
    cluster and the possibly merging group X1. Thus we suggest the
    possibility that its radio properties could be affected by this
    ongoing merger. In particular, if the relic is seen edge-on, and
    if the magnetic field strength is almost uniform in the relic
    region, the observed spectral index trend could be the sign of a
    shock wave moving inward, toward the cluster center. It could
    result from the interaction with X1. Detailed optical and X-ray
    observations would be needed to shed light on this point.}\\
\item{The relic A2345-2 shows the classical feature of ``elongated
  relic sources'' also found in double relics of Abell 3667 and Abell
  3376, as well as in single relic sources as 1253+275 (Andernach et
  al. 1984, Giovannini et al. 1991) and A521 (Ferrari 2003,
  Giacintucci et al. 2008). It is located far from the cluster center,
  its spectral index is steep with mean value $\sim$ 1.3 and steepens
  towards the cluster center, as expected by relic formation theories
  if the relic is observed edge-on. The value of the
  equipartition magnetic field, the direction of the E vectors and the
  detected level of polarization are consistent with previous
  observations of elongated relics and agree with expectations from
  theoretical models as well.  The polarized emission image reveals
  the arc-like structure of the relic A2345-2. If we assume that the
  relic is originated by a spherical shock wave, we can infer the
  propagation center of the shock by extrapolating the curvature
  radius of the relic. It results that the propagation center is $\sim
  2.6'$, southern the present X-ray center of the cluster Abell 2345
  (see Fig. \ref{fig:A23451spixprofile.ps}).\\ This corresponds to a
  physical distance of 450 kpc at this redshift.  From weak lensing
  analysis the galaxy velocity dispersion in this cluster results
  $\sim$900 km/s (Dahle et al. 2002; Cypriano et al. 2004). As we will
  see in Sect.\ref{sec:discussion}, the expected Mach number is of
  about 2.2 for this relic. Since the galaxy velocity dispersion is
  comparable to the sound speed in the ICM (see e.g. Sarazin 1988), a
  Mach number 2.2 corresponds to a velocity of $\sim$2000 km/s.  The
  relic A2345-2 is $\sim$800 kpc far from the spherical-shock
  center.  A shock wave with $M\sim$2.2 travels this distance in
    $\sim$ 0.4 Gyr (if the shock speed remains constant). Thus the
  merging between the two substructures should have occurred at $\sim$
  1200 km/s to explain the shift of the X-ray center in this
  scenario. This is a reasonable value for cluster merger
  velocity.\\ Although a precise estimate should consider the amount
  of energy injected in the ICM as the shock wave passes through it,
  and despite the number of assumptions and approximations, we suggest
  that the relic indicates the position of the merger center as it was
  $\sim$ 0.4 Gy ago. The time that the shock wave has taken to get the
  present relic position is the time that the sub cluster has taken to
  get the current X-ray center position.\\}
\end{itemize}
\begin{figure}[h!]
\centering
   \includegraphics[width=0.9\columnwidth]{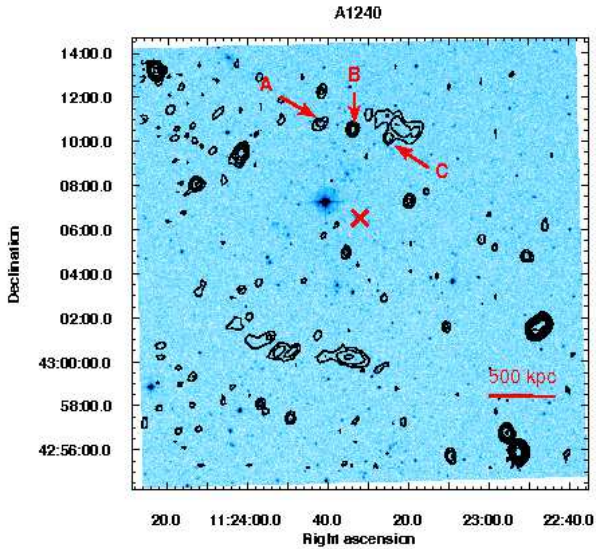}
      \caption{Abell 1240. Colors: Optical emission from DPOSSII (red
      band); Contours: radio emission at 1.4 GHz (HR image). Contours
      start at $\pm$3$\sigma$ and are then spaced by 2. Red cross
      signs the X-ray center, labels refer to the discrete sources
      embedded in A1240-1. }
         \label{fig:A1240_otticoradio.ps}
   \end{figure}

\section{The Cluster Abell 1240}
\begin{figure*}[t!]
\centering
\includegraphics[width=1.99\columnwidth]{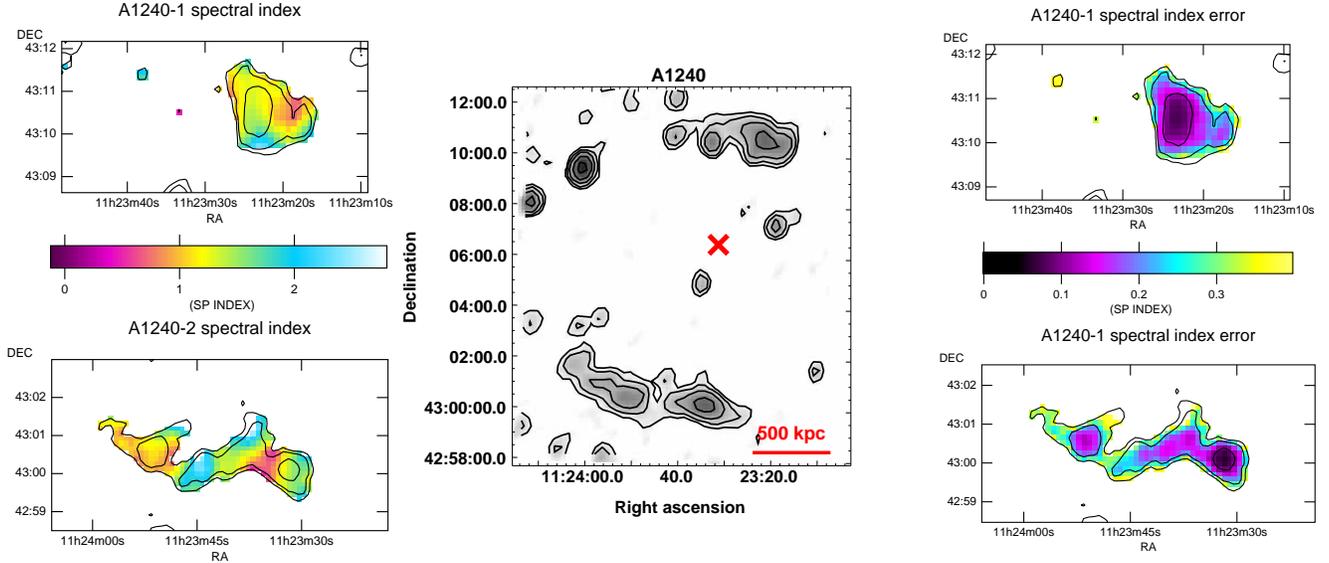}
\caption{Center: the cluster Abell 1240 radio emission at 1.4 GHz. The
beam is 42$''$$\times$33$''$. Contours start at 3$\sigma$ (0.13
mJy/beam) and are then spaced by a factor 2. The cross marks the
cluster X-ray center. Left: colors represent the spectral index of the
relic A1240-1 (top) and A1240-2 (bottom) superimposed over the radio
emission at 325 MHz (contours) The beam is 42$''$$\times$33$''$, first
contours are 2 $\sigma$ (2 mJy/beam), 3$\sigma$ and are then spaced by
a factor 2. Right: Spectral index error image (colors) superimposed
onto the emission at 325 MHz (contours are as above).}
\label{fig:A1240spix}
\end{figure*}

Little is known in the literature about this cluster. It is a rich
cluster classified as Bautz-Morgan type III. In Tab. \ref{tab:x}
general data about this cluster are reported.\\ Kempner \& Sarazin
(2001) have revealed the presence of two roughly symmetric relics from
the Westerbork Northern Sky Survey (WENSS). From WENSS images relics
are visible at 2 and 2.5 $\sigma$ level. Our VLA observation confirm
the presence of two weak radio emitting regions in the cluster's
outskirts. The radio image of the cluster is shown in
Fig. \ref{fig:A1240_otticoradio.ps} (contours) overlaid into optical
emission (from the DPOSSII, red band). The Northern relic (A1240-1) is
located at $\sim$ 270$''$ from the cluster X-ray center. This distance
corresponds to $\sim$ 700 kpc at the cluster's redshift. This relic is
mainly elongated in the E-W direction, and its radio brightness
decreases going from the western to the eastern  part of the relic (see
Fig. \ref{fig:A1240spix}). At 325 MHz only the eastern brightest part
is visible. This is likely due to the higher noise in the 325 MHz
image. In fact from the mean brightness of the weaker part of the
relic, we estimated that it should have a spectral index $>$3 to be
detected at 325 MHz. Three radio sources are embedded in the relic
emission, they are labeled with A B and C in
Fig. \ref{fig:A1240_otticoradio.ps}.  The sources A and B are not
detected in the 325 MHz observations. This is consistent with spectral
index values $<$1, as commonly found in radiogalaxies. A weak emission
at 1.4 GHz links the A radio source at the relic (see
Fig. \ref{fig:A1240spix}). \\
The southern relic (A1240-2) is located at
$\sim$ 400$''$ (1.1 Mpc) from the cluster X ray center. At 1.4 GHz it is
elongated in the E-W direction extending $\sim $ 480$''$. No discrete
sources have been found embedded in the relic emission. Also in this
case at 325 MHz the relic's extension is reduced to $\sim$ 350$''$ along
the main axis, and only the brightest regions are visible at 325
MHz.\\ The relic's physical parameters are reported in
Tab. \ref{tab:A1240_relics}. The quantity are computed excluding the
region where discrete sources (A,B and C) are present.\\

\begin{figure*}[t!]
    \centering
   \includegraphics[width=9cm]{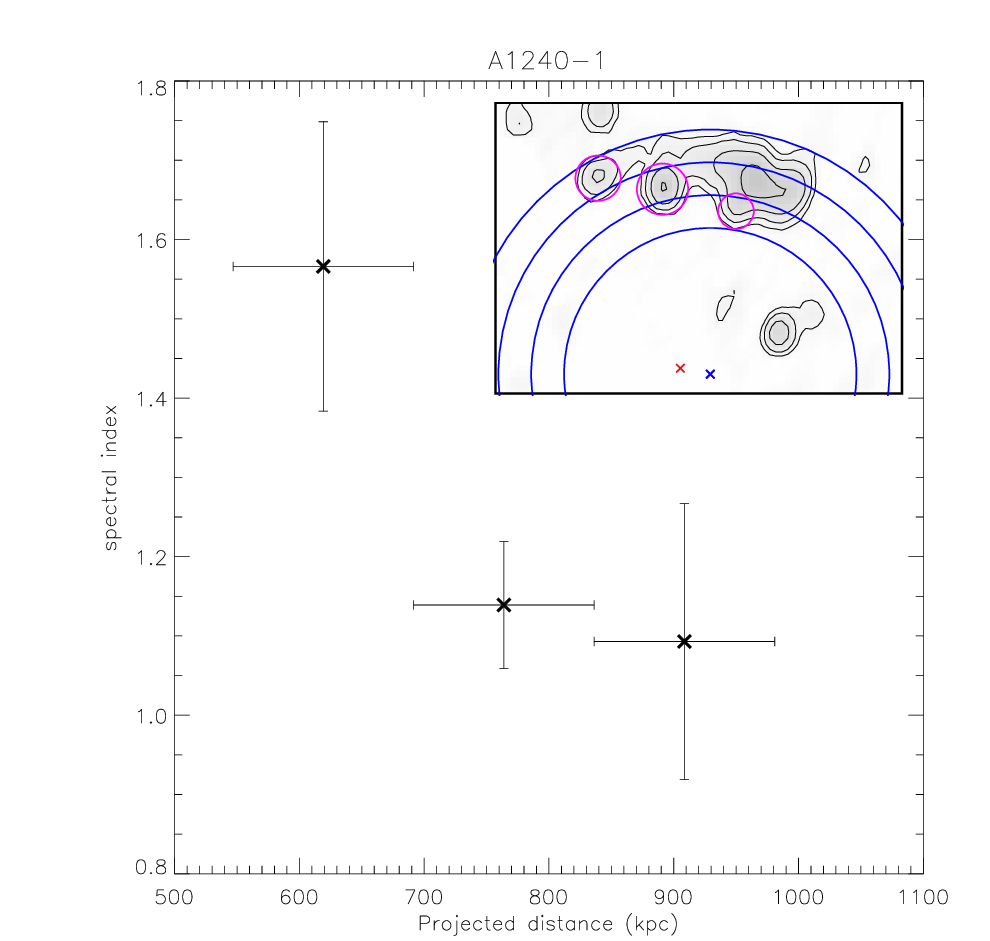}
   \includegraphics[width=9cm]{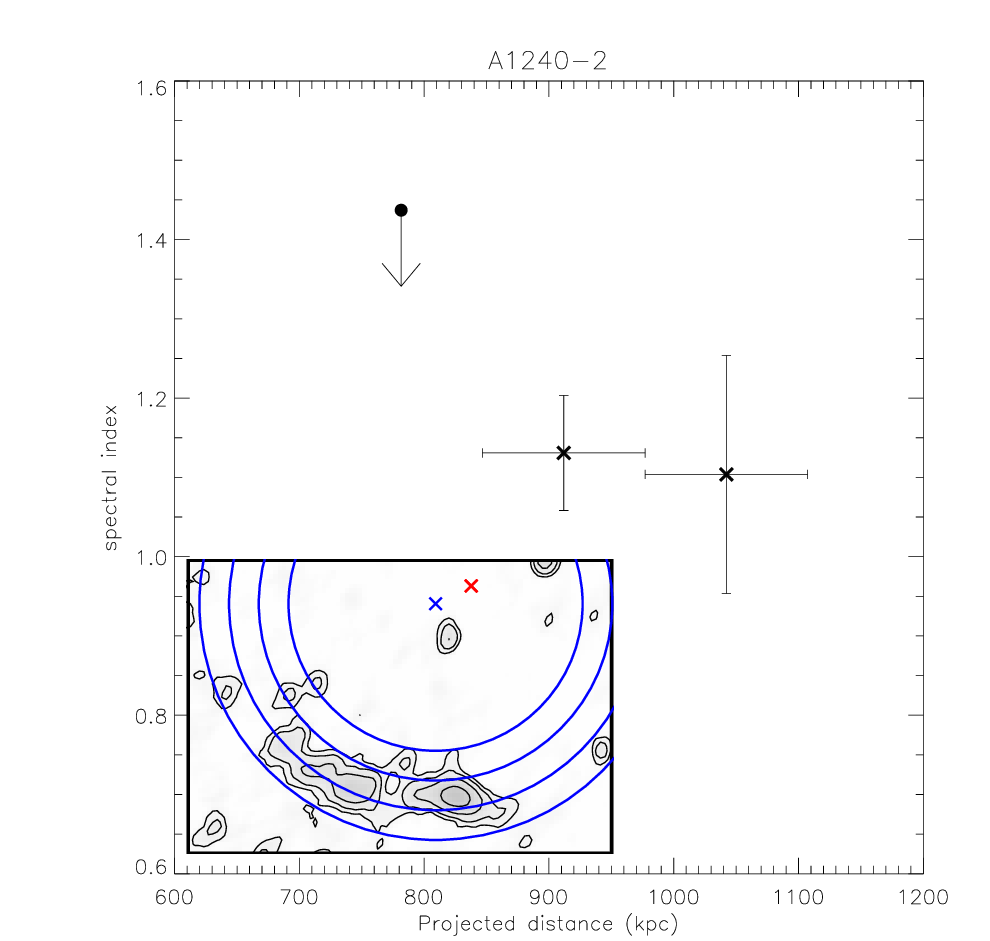}
      \caption{Spectral index radial trend of A1240-1 (left) and
          A1240-2 (right), computed in shells of $\sim 50''$ in
          width. It has been computed excluding the contribution of
          the discrete sources. Crosses refer to spectral index values
          computed in shells where the mean brightness is $>3\sigma$
          at both 325 MHz and 1.4 GHz. Arrows are $3\sigma$ upper
          limits on the spectral index mean value (see text). In the
          inset: displacement of the shells over which the mean
          spectral index has been computed. Circles refer to the
          discrete sources embedded in the relic emission. The red
          cross refers to the cluster X-ray center, the blue cross is
          the center of the spherical shells. }
  \label{fig:A1240spixprofile.ps}
 \end{figure*}

\subsection{Spectral index analysis}
We report in Fig.\ref{fig:A1240spix} the spectral index map and the
spectral index map error for the relics of Abell 1240. They have been
obtained considering only those pixels that have a brightness
$>$2$\sigma$ at both frequencies.\\ Fig. \ref{fig:A1240spix} shows
that the spectral index image is patchy. The spectral index image rms,
$\sigma_{spix}$, is $\sim$ 0.3 and 0.4 for A1240-1 and A1240-2
respectively, while the mean of the spectral index error image,
$<$Spix Noise$>$ is $\sim$ 0.2 for both of the relics. We can then
conclude that features in A1240-2 are statistically significant, while
given the small difference between $\sigma_{spix}$ and $<$Spix
Noise$>$ in A1240-1, we cannot exclude that local features are a noise
artifact in this case. In the relic A1240-2 a gradient is visible
along the main axis of the relic, as has been found in Abell 2256 by
Clarke \& Ensslin (2006).\\ In Fig. \ref{fig:A1240spixprofile.ps} the
radial spectral index trend is shown for A1240-1 and A1240-2
respectively. They have been obtained as described in
Sec. \ref{A2345spectralindex}. Spherical shells are centered close to
the X-ray cluster center and they are parallel to the main axis of
both relics.\\ Despite the small extension of the relics at 325 MHZ,
it is still possible to derive some important results on the spectral
index radial trends in these relics: in the relic A1240-1 the spectral
index is steeper in the inner part of the relic and flatter in the
outer part, as found in A2345-2 and predicted by ``outgoing merger
shock'' models if relics are seen edge-on (Roettiger et al. 1999;
Bagchi et al.  2006). The same trend is consistent with the spectral
index profile derived in A1240-2, although a firm conclusion cannot be
derived from these data. We note in fact that errors and upper limit
in the inner shell cannot exclude a constant spectral index or even an
opposite trend.\\
\begin{figure}[h]
\centering
\includegraphics[width=10cm]{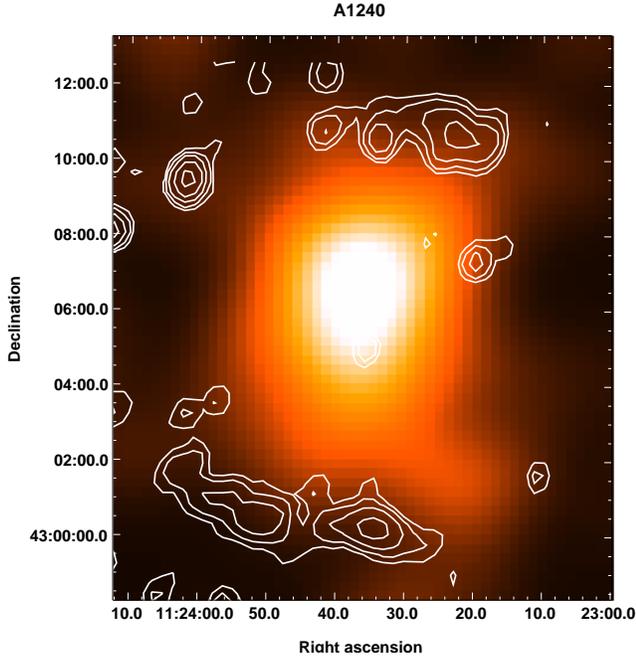}
\caption{Colors: Abell 1240X-ray emission in the energy band 0.5-2 keV
from ROSAT PSPC observations. The image has been smoothed with a
Gaussian of $\sigma\sim$60$''$; contours represent the radio image of
the cluster at 1.4 GHz. The beam is 42$''$$\times$33$''$. First contour is
0.13 mJy/beam, further contours are then spaced by a factor 2. }
\label{fig:A1240X}
\end{figure}  
\subsection{Radio-X-ray comparison}
 We retrieved from the ROSAT data archive X-ray observations in the
energy band $0.5-2$ keV. The cluster is $\sim$ 28$'$ offset from the
center of the ROSAT pointing. Observations have been performed with
the ROSAT PSPC detector for a total exposure time of $\sim$ 12
ksec. After background subtraction the event file has been divided by
the exposure map. We smoothed the resulting image with a Gaussian of
$\sigma=60''$. The resulting image is shown in
Fig. \ref{fig:A1240X}.\\In Fig. \ref{fig:A1240X} the X-ray emission of
the cluster is superimposed onto radio contours. The X-ray emission of
this cluster is elongated in the S-N direction and shows a double
X-ray morphology. As already stated by Kempner \& Sarazin (2001) this
morphology is consistent with a slightly asymmetric merger. \\ Relics
are located at the edge of the X-ray emission. Their emission shows
the characteristic elongated shape, and their main axis is
perpendicular to the main axis of the X-ray emission, as found in
double relics of Abell 3367 and Abell 3376.\\
\begin{figure}[t!]
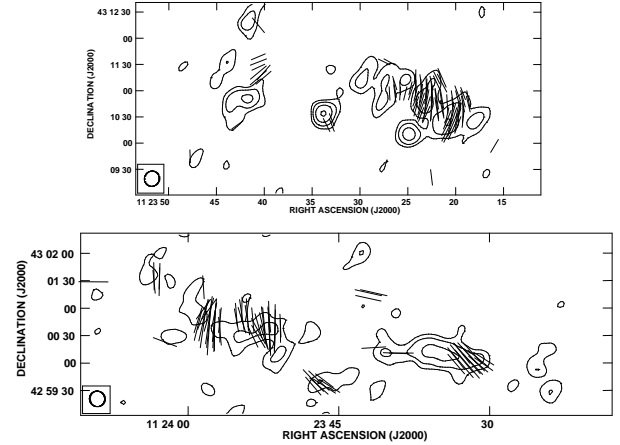

\centering
\includegraphics[width=3cm, angle=-90]{A1240-1pol.ps}
\centering
\includegraphics[width=3cm, angle=-90]{A1240-2pol.ps}
\caption{Top panel: A1240-1 radio emission at 1.4 GHz, lines represent
the E vectors. The line direction indicates the E vector direction and
the line length is proportional to the polarized flux intensity. 1$''$
corresponds to 3$\mu$Jy/beam. The beam is
18$''$$\times$18$''$. Contours start at 0.12 mJy/beam and are then
spaced by a factor 2. Bottom panel: A1240-2 radio emission at 1.4
GHz. The line direction indicates the E vector direction and
the line length is proportional to the polarized flux intensity. 1$''$
corresponds to 2$\mu$Jy/beam. Contours are as above.}
\label{fig:A1240pol}
\end{figure}
\begin{table*} 
\caption{ Abell 1240}
\label{tab:A1240_relics}
\centering
\begin{tabular} {|c  c c c c c c  |} 
\hline\hline
Source name &  Proj. dist & LLS         & F$_{20 cm}$  &  F$_{90 cm}$ & B$_{eq}$ - B$'_{eq}$     & $<\alpha>$ \\ 
            &  kpc        & kpc         & mJy         &  mJy        & $\mu$G     &             \\
\hline
Abell 1240-1 &  270$''$=700 & 240$''$= 650  &6.0$\pm$0.2  & 21.0$\pm$0.8& 1.0 -2.4   & 1.2  $\pm$0.1  \\
Abell 1240-2 &  400$''$=1100& 460$''$= 1250 &10.1$\pm$0.4  & 28.5$\pm$1.1& 1.0 -2.5  & 1.3   $\pm$0.2  \\
\hline \multicolumn{7}{l}{\scriptsize Col. 1: Source name; Col. 2:
projected distance from the X-ray centroid; Col. 3: Largest linear
scale measured on the 20 cm images.}\\ \multicolumn{7}{l}{\scriptsize
Col. 4 and 5: Flux density at 20 and 90 cm; Col. 6: equipartition
magnetic field computed at fixed frequency - fixed energy (see Sec. \ref{sec:A1240pol})
}\\ 
\multicolumn{7}{l}{\scriptsize Col. 7: mean spectral index in
region where both 20 cm and 90cm  surface brightness is $>$ 3 $\sigma$.}
\end{tabular}
\end{table*}

\subsection{Equipartition magnetic field}
\label{sec:A1240pol}
Under the same assumptions explained in Sec. \ref{sec:A2345pol} , we
calculated the equipartition magnetic field for the relics A1240-1 and
A1240-2. Values obtained are reported in
Tab. \ref{tab:A1240_relics}. We note that these values have been
computed considering the brightness of those pixels for which we have
well constrained information about the spectral index value,
i. e. those regions whose emission is detected at both
frequencies. Since the emission at 325 MHz is only detected in a small
region of the relics, while at 1.4 GHz relics are more extended, the
equipartition estimates refer to the same small regions, and different
estimates could be representative of the wider relic emission detected
at 1.4 GHz.\\ We derived the minimum non thermal energy density in the
relic sources from $B'_{eq}$ obtaining $U_{min}\sim$5.5
10$^{-13}$erg/cm$^{-3}$ for A1240-1 and A1240-2. The corresponding
minimum non-thermal pressure in then $\sim$3.4 and $\sim$3.5
10$^{-13}$erg/cm$^{-3}$. The consistency between magnetic field
  equipartition values and magnetic field lower limits derived by
  X-ray emission in other few clusters (see discussion in
  Sec. \ref{secs:Equip}) indicates that equipartition magnetic field
  can be used as a reasonable approximation of the magnetic field in relics.
\subsection{Polarization analysis}
We obtained the polarized intensity images for the relics as described
in Sec. \ref{sec:A2345pol}. In Tab. \ref{tab:radiopol} the parameters
relative to the polarization images of the relics A1240-1 and A1240-2
are reported.\\ In Fig. \ref{fig:A1240pol} the polarized emission of
the two relics is shown. Observations performed with C array cannot
reveal the weak extended emission, and thus only the most compact an
bright regions are visible in this image. In these regions the
magnetic field is mainly aligned along the relic main axis in both of
the relics. This is consistent with what has been observed in the
relics of Abell 2345 and with what is expected from the models that
explain the origin of these sources (e. g. Ensslin et al. 1998;
Roettiger et al. 1999). The mean fractional polarization of A1240-1 is
26\%, reaching values up to 70\%. In the relic A1240-2 the mean
fractional polarization is 29\%, reaching values up to 70\%. From
Eq. \ref{eq:pol} we derive that $B_r^2/B_o^2 \sim$1.4 and 1.2
respectively. Because of possible beam depolarization, internal
depolarization and ICM depolarization, we conclude that
$B_r^2/B_o^2 <$1.4 and $<$1.2. This means that the magnetic energy
density in the random and ordered component is similar.
\subsection{Results on Abell 1240 }
Our observations confirm the presence of two relics in
Abell 1240 with as steep spectral index values as $\sim$1.2 and
$\sim$1.3. The spectral index trends derived for these relics indicate
a radial flattening toward the cluster outskirts. This
is the trend predicted by ``outgoing merger shock'' models.\\ The
double relics radio morphology and location are similar to the double
relics found in Abell 3667 and Abell 3376.\\ The polarization level is
high in both of the relics, although we have to consider that our
polarization observations lack the weak extended regions, that are
probably less polarized. The magnetic field estimate achieved under
the minimum total energy assumption reveals magnetic field of the
order of $\mu$G at the cluster periphery in the relic regions, ordered
on Mpc scale, indicating a magnetic field amplification and ordering.
\section{Discussion}
\label{sec:discussion}
We confirm the presence of double relics in the cluster Abell
1240. Their symmetry and properties strongly suggest a common origin
of A1240-1 and A1240-2.\\ In the cluster Abell 2345 we confirm the
existence of two relics. However, while A2345-2 is a classic extended
peripheral relic source similar to 1253+275, in the Coma cluster (see
Giovannini et al. 1991 and references therein), A2345-1 shows a more
complex structure. We suggest that its properties could be due to its
peculiar position in between the cluster Abell 2345 and the possibly
merging group X1, and thus affected by a more recent merger.\\ Several
models have been proposed to explain the origin of radio relics. They
can be divided into 2 classes:
\begin{enumerate}
\item{Diffusive Shock Acceleration by Fermi-I process (Ensslin et
al. 1998; Roettiger et al. 1999; Hoeft \& Br{\"u}ggen 2007). }
\item{Re-acceleration of emitting particles due to adiabatic
compression of fossil radio plasma (Ensslin \& Gopal-Krishna 2001). }
\end{enumerate} 
In both of these models the presence of a shock within the gas is
required.  The second one also requires the presence of a nearby radio
source to provide the fossil radio plasma which can be re-energized by
the shock wave. Simulations of cluster mergers show indeed that the
merging of two sub-clusters leads to the formation of shocks in the
cluster outskirts (Ryu et al. 2003).\\ In favor of the second scenario
there is the observational evidence that relics resemble individual
objects and do not trace the entire shock front (Hoeft et al.
2004). Moreover, when a radio ghost is passed by a shock wave with
typical velocity of 10$^3$ km/s, it is adiabatically compressed
because of the higher value of the sound speed in the radio ghost
(Ensslin \& Br\"uggen 2002). We have to remind, however, that the
equation of state of the radio emitting plasma is still poorly known,
and that if the radio plasma has a high mass load due to undetectable
cool gas , it should get shocked (Ensslin \& Gopal Krishna 2001).\\
The presence of double relics itself favors the first scenario,
because of the low probability to find two symmetric regions with
fossil radio plasma. \\ Several independent cosmological simulations
have identified two main categories of cosmological shocks:\\ (i)
``accretion shocks'' resulting from accretion of cold gas onto already
formed structure, characterized by high Mach numbers;\\ (ii) ``merging''
or ``internal'' shocks due to merging of substructures such as galaxy
clusters or groups, with moderate Mach numbers: $2\le M \le 4$ (see
review by Bykov et al. 2008 and references therein).
\subsection{Relics from merging shocks}
The presence of double relics is particularly interesting in this
scenario since the shape, morphology and properties of these extended
structures strongly suggest the presence of shock waves propagating
from the cluster center to the peripheral regions. Because of the
short radiative lifetime of relativistic electrons, radio emission is
produced close to the location of the shock waves. These models
predict that the magnetic field is aligned with the shock front and
that the radio spectrum is flatter at the shock edge, where the radio
brightness is expected to decline sharply.  \\ The shock compression
ratio can be estimated from the radio spectral index $\alpha$ (assuming an
equilibrium electron population accelerated and cooled at the same
time, and assuming a polytropic index 5/3, see Drury 1983), as
\begin{equation}
R=\frac{\alpha+1}{\alpha-0.5}.
\end{equation}
The pressure and temperature jumps across the shock can be estimated
 from the theory of shocks (Landau \& Lifschitz 1966) as
\begin{equation}
\frac{P_2}{P_1}=\frac{4R-1}{4-R}= \frac{\alpha+1.5}{\alpha-1}; \frac{T_2}{T_1}=\frac{P_2}{RP_1}
\end{equation}
here and after the index 2 refers to down stream regions and 1 to
 up-stream regions i.e. regions inside and outside the cluster shock
 front. These parameters are reported in Tab. \ref{tab:DSA}.
\begin{table*}
\caption{Predictions from the shock acceleration model}
\label{tab:DSA}
\centering
\begin{tabular}{|c c c c c c c |} 
\hline\hline
Relic        &$\alpha$ & M &  R    & $P_2/P_1$    &  $T_2/T_1$ & $(B_2/B_1)_{isoP}$\\
\hline     
Abell 2345-1 &1.5$\pm$0.1 &2.8$\pm$0.1&2.5$\pm$0.2&6$\pm$1  & 2.4$\pm$0.4  &2.4$\pm$0.2\\
Abell 2345-2 &1.3$\pm$0.1 &2.2$\pm$0.1&2.9$\pm$0.2&9$\pm$3  & 3$\pm$1      &3.0$\pm$0.5\\
Abell 1240-1 &1.2$\pm$0.1 &3.3$\pm$0.2&3.1$\pm$0.3&14$\pm$6 & 4$\pm$2      &3.7$\pm$0.8\\
Abell 1240-2 &1.3$\pm$0.2 &2.8$\pm$0.3&2.9$\pm$0.4&9$\pm$3  & 3$\pm$2      &3.0$\pm$0.5\\
\hline
\multicolumn{7}{l}{\scriptsize Col. 1: 
Source name; Col 2: spectral index value; Col. 3:  Mach number; Col 4: Shock compression ratio estimated   }\\
\multicolumn{7}{l}{\scriptsize from the radio spectral index;  Col. 5, 6: 
Pressure and temperature jump across the shock; Col. 7: Magnetic field  }\\
\multicolumn{7}{l}{\scriptsize strength in the pre and post shock regions required to
support the relic against the thermal pressure }\\
\end{tabular}
\end{table*}
The Mach number of the shock can be estimated from the radio spectral
index under some assumptions: if the emitting particles are linearly
accelerated by shock, the spectral index of the particle energy
spectrum p ($=2\alpha+1$) is related to the Mach number $M$ of the
shock through:
\begin{equation}
p=2\frac{M^2+1}{M^2-1}+1
\end{equation}
including the effect of particle aging (continuous injection and
Inverse Compton energy losses, see e.g. Sarazin 1999). Mach number
values we obtained are reported in Tab. \ref{tab:DSA}. These values
are lower than Mach number expected for accretion shocks (e. g. Bykov
et al. 2008), and are instead consistent with those expected for
weaker shocks due to merging of structures. \\ The spectral index
trend clearly detected in A2345-2 and in both relics of Abell 1240
agrees with the predictions of this scenario. If relics are seen
edge-on, the flattest region, in the outer part of the relics, would
corresponds to the current shock location, indicating shock waves
moving outward from the cluster center. As discussed
  in Sec.  \ref{Sec:A2345results}, A2345-1 shows a more complex radio
  emission. It could be affected by a more recent merger with the X1
  group. It could trace a merger shock moving inward to the cluster
  center as a result of the Abell 2345 - X1 group interaction.\\
\subsubsection{Magnetic field and merging shocks}
\label{sec:Magnetic-Shock}
 The study of the magnetic field associated with the relics offers
 further opportunities to investigate the connection between relics
 and merger shock waves. First of all the presence of relics itself
 indicates the existence of significant magnetic field at the cluster
 periphery on the Mpc scale. Furthermore, the detected level of
 polarization shows that the magnetic field in these regions is rather
 ordered. \\ The effect of passage of a shock wave in the ICM
   could be twofold: (i) order and compress a magnetic field that was
   randomly oriented before the shock passage or (ii) compress a
   magnetic field that was already ordered on the relic scale before
   the shock passage . This depends on the turbulence development at
   the cluster periphery, that could either give rise to a random
   field in the cluster outskirts (case i) or not (case ii). Little is
   known about this point from observational point of
   view. Observational evidence from the gas pressure map of the Coma
   cluster (Schuecker et al. 2004) indicates the relevance of chaotic
   motions within the ICM.  Cosmological numerical simulations
   (e.g. Bryan \& Norman 1998; Sunyaev, Bryan \& Norman 2003) suggest
   that the level of ICM turbulence is larger at increasing radial
   distances from the cluster center. If the simple
   Kolmogorov's picture of incompressible fluid turbulence is assumed, this
   implies a more developed turbulence in the outermost region (since
   the decay time is L/$\sigma$, where L is the typical scale where
   the bulk of turbulence is injected, and $\sigma$ is the rms
   velocity of turbulence). Recently Ryu et al. (2008) argued that
   turbulence is likely well developed in clusters and
   filaments, and not in more rarefied regions such as sheets and
   voids. On the other hand Dolag et al.(2005) suggested that the bulk
   of turbulence is injected in the core of galaxy clusters, thus
   implying a more developed turbulence in the innermost regions,
   compared to the outermost ones.  The main limitation of
   cosmological simulations is the lack of resolutions in low density
   environments, that makes it difficult to discriminate if the
   turbulent cascade is developed in these regions. Moreover, details
   of the conversion process of large–scale velocity fields into MHD
   modes is still poorly understood. Thus, from the theoretical point
   of view the overall picture seems still uncertain.\\
% However, present-day numerical simulations
%are just facing the necessary space resolution to completely follow the
%injection and cascade (if any) of turbulent eddies in the ICM during
%cluster mergers and/or accretions, and thus from the theoretical point of
%view the overall picture seem still uncertain.
%Indeed, the magnetic
 %  field in the cluster center is expected to be isotropic and with
 %  random orientation. (see e. g. Govoni et al. 2006; Murgia et
  % al. 2004; Guidetti et al. 2008).  From cosmological simulations
 %  From cosmological simulations Bryan \& Norman (1998) and Norman \&
 %  Bryan (1999) find that within the virial radius the ICM is in a
  % turbulent state with eddies ranging from tens to hundreds kpc in
 %  size; Ryu et al. (2008) argue that it is likely that turbulence is
 %  well developed in clusters and filaments, and not in more rarefied
 %  regions such as sheets and voids. Moreover, according to Dolag et
  % al. (2005) details of the conversion process of large–scale
  % velocity fields into MHD modes is still poorly understood. The main
  % limitation of cosmological simulations is the lack of resolutions
  % in low density environments, that makes it difficult to
  % discriminate if the turbulent cascade is developed in these
  % regions.\\ 
In the case that the magnetic field in the cluster
outskirts is randomly oriented before the shock passage (i. e. the
turbulence is developed in the cluster outskirts) and that it has been
amplified and ordered, by the passage of the shock wave (case i
above), the observed ratio $B_r/B_o$ derived by polarization analysis
(Sec. \ref{sec:A2345pol} and \ref{sec:A1240pol}) could be used to
estimate the magnetic field amplification due to the passage of the
shock.\\
%Indeed,
% the role of shock waves in the ICM could be fundamental in order to
% generate or amplify the intergalactic magnetic field (Kulsrud et
% al. 1997; Ryu et al. 1998). \\ 
 Following Ensslin et al. (1998), if the relic is seen at some angle
 $\delta>$0 between the line of sight and the normal of the shock
 front, the projected magnetic field should appear perpendicular to
 the line connecting the cluster center and the relic. This is indeed
 what polarization data presented here show. The magnetic field
 amplification, the observed integral polarization and the
 preferential direction of the field revealed by the E vectors
 orientation could be derived, provided that $\delta$ and R, the shock
 compression factor, are known. Present data do not allow to infer the
 angle $\delta$. Future X-ray and optical observations could
 reconstruct the merging geometry for these two clusters, as done,
 e.g.  in Abell 521 by Ferrari et al. (2003, 2006). Despite this, if
 relics are supported by magnetic pressure only, the upstream and
 downstream fields are related by $(B_2^2/B_1^2)_{isoP}=P_2/P_1$
 (``strong field'' case in Ensslin et al. 1998). This ratio can be
 compared to the ratio derived by the polarization properties of the
 relics, under the assumption that $B_2$ corresponds to the ordered
 component of the field and $B_1$ to the random one. In
 Tab. \ref{tab:DSA} the $(B_2/B_1)_{isoP}$ ratio is reported for the
 relics in Abell 2345 and Abell 1240. These values are comparable to
 the observed ratio $B_r/B_o$ derived by polarization analysis
 (Sec. \ref{sec:A2345pol} and \ref{sec:A1240pol}).  \\ Another
   indication of the magnetic field amplification in the relics may be
   obtained by comparing the magnetic field in the relic with the
   cluster magnetic field intensity expected at the relic
   location. Relics are located at 700-1100 kpc from the cluster
   center in Abell 2345 and Abell 1240. At these distances the cluster
   magnetic field strength is expected to be of the order of
   $\sim$10$^{-1}\mu$G (see e. g. Dolag et al. 2008; Ferrari et
   al. 2008 and references therein).  Equipartition magnetic field
   values are of the order of $\mu$G (see Sec. \ref{secs:Equip} and
   \ref{sec:A1240pol}), thus about 10 times higher. Despite the number
   of uncertainties and assumptions related with the equipartition
   estimate, this is consistent with the ratio $(B_2^2/B_1^2)_{isoP}$
   and $B_r/B_o$.\\Even if no firm conclusion can be obtained by this
   analysis, we can conclude at least that the drawn picture has no
   inconsistencies with the presented observations.
\subsection{Relics from Adiabatic compression}
Another model to explain the origin of cluster radio relics has been
proposed by Ensslin \& Gopal-Krishna (2001). This idea has been
investigated with the help of 3-dimensional Magneto Hydro Dynamical
simulations by Ensslin \& Br\"uggen (2002) and in a more realistic
cosmological environment by Hoeft et al. (2004). In this scenario
cluster radio relics would originate by the compression of fossil
radio plasma by shock wave occurring in the process of large scale
structure formation. The expected high sound velocity of that still
relativistic plasma should forbid the shock to penetrate into the
radio plasma, so that shock acceleration is not expected in this
model. The plasma gains energy adiabatically from the compression and
the magnetic field itself is amplified by such compression. If the
electron plasma is not older than 2 Gyr in the outskirts of a cluster,
they can emit radio wave again.  Simulations performed by Ensslin \&
Br\"uggen (2002) show that the radio morphology of the resulting radio
relic in the early stage after the shock passage is sheet-like. Then
the formation of a torus is expected when the post shock gas starts to
expand into the volume occupied by the radio plasma. Thus it is
expected in this scenario that some correlation should exist between
the morphology of the radio relic and its spectral index, that traces
the time passed after the shock wave has compressed and re-energized
the emitting particles. A2345-1 shows indeed a torus-like radio
structure and a spectral index higher than A2345-2, A1240-1 and
A1240-2, that exhibit a sheet-like structure.  The simulations
performed by Ensslin \& Br\"uggen (2002) indicate that the compression
of the radio plasma by the shock can be estimated from a cluster radio
relic with a toroidal shape. Assuming the idealized case of a
initially spherical and finally toroidal radio cocoon, the compression
factor is given by:
\begin{equation}
\label{eq:plasma}
R'=\frac{2r_{max}^2}{3\pi r_{min}^2}
\end{equation}
where $r_{max}$ and $r_{min}$ refer to the outer and inner radius of
the torus. In the case of A2345-1 we assume that the observed torus
like structure can be described by taking $r_{max}\sim$ the LLS of the
relic and $r_{min}$ the thickness of the filament in the N-E part of
the relic, as suggested by the same authors in the case of non perfect
toroidal filamentary relics. With $r_{max}\sim$ 1 Mpc, $r_{min}\sim$
200 kpc it results $R'\sim$ 5. This is higher than the value of
  the maximum compression ratio for mono-atomic gas (that is 4); this
  would indicate that the radio plasma has a different equation of
  state. However no conclusion can be drawn since Eq. \ref{eq:plasma}
  is based on too simplistic assumptions, in particular a spherical
  model for the compressed relic.\\
%thus we
%conclude that this scenario requires the radio plasma to have a
%different equation of state (for instance if the radio plasma behaves
%like a ultra-relativistic gas then the limit value for R$'$ would be
%7) and the shock wave to be strong to explain such a high compression
%ratio.\\
\section{Conclusions}
We have presented 1.4 GHz and 325 MHz observations of Abell 2345 and
Abell 1240. The presence of double relics in these cluster had been
inferred by Giovannini et al. (1999) for Abell 2345 and by Kempner \&
Sarazin (2001) for Abell 1240 from NVSS and WENSS. We confirm the presence
of two relics in each of these clusters.
By combining 1.4 GHz and 325 MHz observations we obtained the spectral
index image of the diffuse radio emission. The study of the polarized
emission at 1.4 GHz has been presented as well.  The analysis of both
the spectral index distribution and the polarization properties of
relics allows to test several independent predictions of the relic
formation models.\\ We report the summary of the results from the
presented analysis:\\
\begin{enumerate}
\item{ {\bf A2345}: two relics have been detected in the cluster
outskirts at both 1.4 GHz and 325 MHz. They are not perfectly
symmetrical with respect to the cluster center; the normals to the
relic main axis form an angle of $\sim$150$^{\circ}$. A2345-2 is a
classical peripheral relic and A2345-1 is a peculiar relic with a
torus-like structure possibly related to a merging region.}\\
\item{{\bf A1240}: relics are fainter than relics in A2345. Their
extended emission is detected at 1.4 Ghz while only their brightest
part are detected at 325 MHz. They are symmetrical with respect to the
cluster center and the angle between their normals is
$\sim$180$^{\circ}$ as found in the other known double relics: Abell 3667,
Abell 3376 and RXCJ1314.4-2515.}\\
\item{ Relics are located at the edge of the X-ray emission of Abell 2345
and Abell 1240. The X-ray emission of Abell 2345 shows multiple substructures
that could be galaxy groups interacting with A2345. Peculiar features
of A2345-1 could arise from this multiple interaction, but only
detailed X-ray and optical analysis could shed light on this
point.}\\
\item{Relics in Abell 1240 are located perpendicular to the cluster main
axis revealed by X-ray observations. The double X-ray morphology of
the cluster is typical of merging clusters. }\\
\item {The average spectral indexes are steep. We found 1.5 $\pm$ 0.1
  and 1.3 $\pm$0.1 for A2345-1 and A2345-2 and 1.2$\pm$ 0.1, 1.3$\pm$
  0.2 for A1240-1 and A1240-2.}\\
\item{The spectral index distribution in the relics is rather
irregular and patchy, although this, a clear radial trend is present
in the relics of these two clusters.  A2345-2 spectral index ranges
from $\sim$1.5 in the region closer to the cluster center to $\sim$1.1
in the outer rim. This trend is consistent with shock models
predictions. The same trend is observed in both of Abell 1240
relics. A1240-1 spectral index ranges from $\sim$1.1 to $\sim$1.6
going from the outer to the inner rim, A1240-2 spectral index is also
consistent with a similar trend (going from $\alpha <$ 1.5 in the
inner rim to $\alpha\sim$ 1.1 in the outer one).  An opposite trend is
instead detected in A2345-1. Spectral index values are lower in the
inner rim ($\sim$1.3) and increase toward the outer part of the relic
reaching values $\sim$1.7. This trend could be due to its peculiar
position between two merging clumps.}\\
\item{The magnetic field, as revealed by polarized emission is mainly
  aligned with the relic main axis. In Abell 2345 the polarized
  emission reveals the arc-like structure morphology of the relic
  A2345-2. Under equipartition conditions, values of $\sim$ 2.2 - 2.9 $\mu$G are
  derived. The field has been likely amplified,
  consistently with shock models predictions.}\\
\end{enumerate}
 These results have been discussed in the framework of relic
  formation models. The Mach numbers derived from the value of radio
  spectral index disfavour the ``accretion shock'' scenario, since
  they are too small. Outgoing merger shock waves, proposed to explain
  double relic emission in Abell 3667 and A3376, could also work in
  Abell 1240 and Abell 2345. For the latter cluster we suggest that
  the peculiar emission of A2345-1 could be explained by a shock wave
  moving inward, due to the interaction of the main cluster with the
  X1 group.\\ The toroidal shape of A2345-1 could be produced by
  adiabatic compression, however the available data and models do not
  allow a conclusive comparison.
 
\begin{acknowledgements}
    The authors grateful to Franco Vazza, Marica Branchesi and
    Elisabetta Liuzzo for helpful discussions and to Klaus Dolag for
    elucidating discussion on cluster magnetic fields. The authors
    thank the anonymous referee for useful suggestions and
    comments. NRAO is a facility of the National Science Foundation,
    operated under cooperative agreement by Associated Universities,
    Inc. This work was partly supported by the Italian Space Agency
    (ASI), contract I/088/06/0, by the Italian Ministry for University
    and Research (MIUR) and by the Italian National Institute for
    Astrophysics (INAF). This research has made use of the NASA/IPAC
    Extragalactic Data Base (NED) which is operated by the JPL,
    California Institute of Technology, under contract with the
    National Aeronautics and Space Administration.
\end{acknowledgements}


\begin{thebibliography}{}

\bibitem[2006]{Anderbach} Andernach H., Feretti L., \& Giovannini G., 1984, A\&A, 133, 252
\bibitem[1990]{Baars} Baars, Jacob W.M. \&  Martin, Robert M. 1990, LIACo, 29, 293
\bibitem[2006]{Bagchi} Bagchi, J., Durret, F., Lima Neto, G.B., \& Paul, S. 2006, Science, 314, 791  
\bibitem[2005] {BeckKrau} Beck, R. \& Krause, M. 2005, AN, 326, 414
\bibitem[2004]{Boehringer} B\"ohringer, H. et al. 2004, A\&A, 425, 367
\bibitem[2008]{bykov} Bykov, A.M., Dolag, K., \& Durret, F. 2008, SSRv., 134, 119
\bibitem[1997]{Brunetti} Brunetti, G., Setti, G., \& Comastri, A. 1997, A\&A, 325, 898

\bibitem[1998]{BryNor98} Bryan G.L. \& Norman M.L. 1998, ApJ, 495, 80

\bibitem[1966]{Burn} Burn, B.J. 1966, MNRAS 133 67
%\bibitem[1976]{Cavaliere} Cavaliere, A., Fusco-Femiano, R. 1976, A\&A, 49, 137
\bibitem[2008]{Chenetal} Chen, C.M.H., Harris, D.E., Harrison, F.A., \& Mao, P.H. 2008, MNRAS, 383, 1259

%\bibitem[2003]{Churazov}Churazov E., Forman W., Jones C., B{\"o}hringer H. 2003, ApJ, 590, 225


\bibitem[2006] {Clarke} Clarke, T.E., \& Ensslin, T.A. 2006, AJ, 131, 2900
\bibitem [2004]{Cypriano} Cypriano, E.S., Sodr\'e, L.J., Kneib, J.P., Campusano, L.E. 2004, ApJ 613 95
\bibitem[2002]{Dahle} Dahle, H., Kaiser, N., Irgens, R.J., Lilje, P.B., \& Maddox, S.J. 2002 ApJ 139 313
\bibitem[1999]{David} David, L.P., Forman, W., \& Jones, C. 1999, ApJ, 519, 533
\bibitem[2008]{Dolag08} Dolag, K., Bykov, A.M., \&  Diaferio, A. 2008, SSRv, 134, 311
\bibitem[2001]{Dolag01} Dolag, K., Schindler, S., Govoni, F., \& Feretti, L. 2001, A\&A, 378, 777

\bibitem[2005]{Dolag05} Dolag K., Vazza F., Brunetti G., \& Tormen G. 2005, MNRAS, 364, 753

\bibitem[1983]{Drury} Drury, R.O. 1983, Reports on Progress in Physics, 46, 973
\bibitem[1998]{Ensslin08} Ensslin, T.A., Biermann, P.L., Klein, U., \& Kohle, S. 1998, A\&A, 332, 395
\bibitem[2002]{EnssBrug02} Ensslin, T.A., \& Br\"uggen, M. 2002, MNRAS, 331, 1011
\bibitem[2001]{Ensslin01} Ensslin, T.A., \& Gopal-Krishna 2001, A\&A, 366, 26
\bibitem[2006]{FerNei06} Feretti, L., \& Neumann, D.M. 2006, A\&A, 450, L21
\bibitem[2005]{Feretti05} Feretti, L., Schuecker, P., B\"ohringer, H., Govoni, F., \& Giovannini, G. 2005, A\&A 444 157
\bibitem[2003]{Ferrari03} Ferrari, C., 2003, ``Multi-wavelength
  analysis of merging galaxy clusters'', PhD Thesis, available
  electronically at http://tel.archives-ouvertes.fr/docs/00/04/85/55/PDF/tel-00010416.pdf
\bibitem[2003]{Ferrarial03} Ferrari, C., Maurogordato, S., Cappi, A., \& Benoist, C., 2003, A\&A, 399, 813
\bibitem[2003]{Ferrarial06} Ferrari, C., Arnaud, M., Ettori, S., Maurogordato, S., \& Rho, J., 2006, A\&A, 446, 417
\bibitem[2008]{Ferrari08} Ferrari, C., Govoni, F., Schindler, S., Bykov, A.M., \&  Rephaeli, Y. 2008, SSRv, 134, 93
\bibitem[2008]{Giacintucci} Giacintucci, S., Venturi, T., Macario, G., et al. 2008, A\&A in press, 2008arXiv0803.4127G
\bibitem[1991]{Giovannini91} Giovannini, G., Feretti, L., \& Stanghellini, C. 1991, A\&A, 252, 528
\bibitem[2004]{GiovFer04} Giovannini, G., \& Feretti, L. 2004,
Journal of the Korean Astronomical Society, Proceedings of the 3rd
Korean Astrophysics Workshop ``Cosmic Rays and Magnetic Fields in
Large Scale Structure'', Pusan, Korea, August 2004, eds. H. Kang \&
D. Ryu, 37, 1
\bibitem[1999]{Giovannini99} Giovannini, G., Tordi, M., \& Feretti, L. 1999, New Astron., 4, 141
\bibitem[2004]{GovFer04} Govoni, F., \& Feretti, L. 2004, IJMPD, 13, 1549 
%\bibitem[2005]{Govoni06} Govoni, F., Murgia, M., Feretti, L., et al. 2006, A\&A, 460, 425
%\bibitem[2008]{Guidetti} Guidetti, D., Murgia, M., Govoni, F., et al. 2008, A\&A, 483, 699
\bibitem[2001]{Henr&Muk} Henriksen, M., \& Mukshotzky, R. 2001, ApJ, 553, 84.
\bibitem[2004]{Hoeft04} Hoeft, M., Br\"uggen, M., \& Yepes, G. 2004, MNRAS, 347, 389
\bibitem[2007]{Hoeft07} Hoeft, M., \& Br\"uggen, M. 2007, MNRAS, 375, 77
\bibitem[2002]{JohnstonHollitt} Johnston-Hollitt, M., Clay, R.W.,
Ekers, R.D., Wieringa, M.H., \& Hunstead, R. W. 2002 IAUS 199, 157
\bibitem[2001]{Kempner01} Kempner, J.C., \& Sarazin, C.L. 2001, ApJ, 548, 639
\bibitem[2004]{Kempner04} Kempner, J.C., et al. 2004, in ``The riddle of cooling fows in
  galaxies and clusters of galaxies'', T. Reiprich, J. Kempner \& N. Soker (eds.), published
  electronically at http://www.astro.virginia.edu/coolingflow/
%\bibitem[1997]{Kulsrud} Kulsrud, R. M., Cen, R., Ostriker, J. P., \& Ryu, D. 1997, ApJ, 480, 481
\bibitem[1966]{Landau} Landau L.,D. \& Lifshitz, E.M. 1966, Fluid Mechanics, (Pergamon Press Ltd.)
%\bibitem[2004]{Murgia} Murgia M., Govoni, F., Feretti, L., et al. 2004, A\&A, 424, 429
%\bibitem[1999]{NorBria99} Norman M.L, \& Bryan G.L. 1999, LNP Vol. 530:The Radio Galaxy Messier 87, 106
\bibitem[2007]{Orru} Orr\'u, E., Murgia, M., Feretti, L., et al. 2007, A\&A, 467, 943
%\bibitem[2004]{Ota} Ota, N., \& Mitsuda, K. 2004, A\&A, 428, 757
\bibitem[2004]{Owen04} Owen, F.N., Brogan, C.L., \& Clarke T.E. 2004, published
  electronically at http://www.vla.nrao.edu/astro/giudes/p-band/p-reduction/ 
\bibitem[1970]{Pacholczyk} Pacholczyk, A.G. 1970, Radio astrophysics, (Freeman Eds.)
\bibitem[2008]{Pizzo} Pizzo, R.,F.,de Bruyn, A.,G., Feretti., L., \& Govoni, F. 2008, A\&A, 481L 91
\bibitem[1999]{Roettiger} Roettiger, K., Burns, J.O., \& Stone, J.M. 1999, ApJ, 518, 603
\bibitem[1997]{Rottgering} R{\"o}ttgering, H.J.A., Wieringa, M.H.,
Hunstead, R.W., \& Ekers, R.D. 1997, MNRAS, 290, 577
%\bibitem[1998]{Ryu98} Ryu, D., Kang, H., \& Biermann, P.L., 1998, A\&A, 335, 19
\bibitem[2003]{Ryu03} Ryu, D., Kang, H., Hallman, E., \& Jones, T. W. 2003, ApJ, 593, 599        
\bibitem[2008]{Ryu08} Ryu D., Kang H., Cho J., \& Das S. 2008, Science, 320, 909
\bibitem[1988]{Sarazin} Sarazin, C.L. 1988, X-ray emission from clusters of galaxies, (Cambridge University Press)
\bibitem[1999]{Sarazin} Sarazin, C.L. 1999, ApJ, 520, 529
\bibitem[2004] {Schuecker} Schuecker P., Finoguenov A., Miniati F., B{\"o}hringer H., \& Briel
 U.G. 2004, ApP, 426, 387
\bibitem[2008]{Solovyeva} Solovyeva, L., Anokhin, S., Feretti, L., et al. 2008, A\&A 484 621
\bibitem[2003]{Suyaenev} Sunyaev, R. A., Norman, M. L., \& Bryan, G. L. 2003, AstL 29, 783
\bibitem[2007]{Venturi} Venturi, T., Giacintucci, S., Brunetti, G., et al. 2007, A\&A 463 937
\end{thebibliography}
\end{document}